\documentclass[numberedappendix]{emulateapj}

\usepackage{graphicx}
\usepackage{amsmath}
\usepackage{xcolor}
\usepackage{float}
\usepackage{placeins}
\usepackage{multirow}
\usepackage{array}
\usepackage[normalem]{ulem}
\usepackage{tabularx}
\usepackage{threeparttable}
\usepackage{booktabs}
\usepackage{longtable}
\usepackage{hyperref}
\usepackage[title]{appendix}
\usepackage{lipsum}
\newcommand\Bstrut{\rule[-2.5ex]{0pt}{0pt}}
\newcommand\bstrut{\rule[-0.5ex]{0pt}{0pt}}
\newcommand{\remnant}{0509$-$67.5 }
\newcommand{\remnosp}{0509$-$67.5}
\newcommand{\remtwo}{0519$-$69.0}

\newcommand{\ha}{H$\alpha$ }
\newcommand{\chandra}{{\em Chandra}}
\newcommand{\HST}{{\em HST }}

\newcommand{\Lagr}{\mathcal{L}}

\newcommand{\scale}{0.05$^{\prime\prime}\,$pixel$^{-1}$}

\usepackage{hyperref} 
\begin{document}

\title{A Hydro-Based MCMC Analysis of SNR \remnant: Revealing the Explosion Properties from Fluid Discontinuities Alone}

% \author[0000-0002-6688-3307]{Prasiddha Arunachalam}
% \affiliation{Department of Physics and Astronomy, Rutgers University, 136 Frelinghuysen Road, Piscataway, NJ 08854, USA}
% \author[0000-0002-8816-6800]{John P. Hughes}
% \affiliation{Department of Physics and Astronomy, Rutgers University, 136 Frelinghuysen Road, Piscataway, NJ 08854, USA}
% \author[0000-0003-4340-1309]{Luke Hovey}
% \affiliation{Stewardship Data Analysis, Mission Support and Test Services, Los Alamos NM 87544, USA}
% \affiliation{Theoretical Design Division, Los Alamos National Laboratory, Los Alamos NM 87545, USA}
% \author[0000-0001-9567-0880]{Kristoffer Eriksen}
% \affiliation{Theoretical Design Division, Los Alamos National Laboratory, Los Alamos NM 87545, USA}

\title{A Hydro-Based MCMC analysis of \remnant: Revealing the explosion properties from fluid discontinuities alone}

% \title{No Gas, no mess: Measuring the explosion properties of SNR \remnant using fluid discontinuities }

% \title{The shocking past: A data-driven approach to measuring the explosion properties of SNR \remnant using fluid discontinuities }

\author{Prasiddha Arunachalam$^{1}$ John P. Hughes$^{1}$, Luke Hovey$^{2}$, and Kristoffer Eriksen$^3$}
\affil{$^1$Department of Physics and Astronomy, Rutgers University, 136 Frelinghuysen Road, Piscataway, NJ 08854, USA\\
$^2$Stewardship Data Analysis, Mission Support and Test Services, Los Alamos NM 87544, USA\\
$^3$Theoretical Design Division, Los Alamos National Laboratory, Los Alamos NM 87545, USA}

\begin{abstract}
     Using \textit{HST} narrow-band H$\alpha$ images of supernova remnant 0509$-$67.5 taken $\sim$10 years apart, we measure the forward shock (FS) proper motions (PMs) at 231 rim locations. The average shock radius and velocity are $3.66 \pm 0.036$ pc and $6315 \pm 310$ km s$^{-1}$. Hydrodynamic simulations, recast as similarity solutions, provide models for the SNR's expansion into a uniform ambient medium. These are coupled to a Markov chain Monte Carlo (MCMC) analysis to determine explosion parameters, constrained by the FS measurements. For baseline explosion parameters, the MCMC posterior distributions yield an age of $315.5 \pm 1.8$ yr, a dynamical explosion center at $5^{\rm h}09^{\rm m}31^{\rm s}.16$, $-67^{\circ}31^{\prime}17.1^{\prime\prime}$ and ambient medium densities at each azimuth ranging over $3.7-8.0 \times 10^{-25}$ g cm$^{-3}$. 
     We can detect stellar PMs corresponding to speeds in the LMC of 770 km s$^{-1}$ or more using the H$\alpha$ images. Five stars in the remnant show measurable PMs but none appear to be moving radially from the center, including a prominent red star $4.6^{\prime\prime}$ from the center.
     Within a 1.4$^{\prime\prime}$ radius of our dynamical center, there are 4 stars, including  3 faint, previously unidentified ones.
     Using coronal [Fe XIV] $\lambda$5303 emission as a proxy for the reverse shock location, we constrain the explosion energy (for a compression factor of 4) to a value of $E=(1.30\pm0.41)\times10^{51}\, \text{erg}$ for the first time from shock kinematics alone. Higher compression factors (7 or more) are strongly disfavored based on multiple criteria, arguing for inefficient particle acceleration in the Balmer shocks of \remnosp.
     
     % We consider two different FS compression factors (4 and 7) to approximate the effects of particle acceleration. The higher compression factor is rejected because it requires an unphysically large explosion energy, providing further evidence that cosmic-ray acceleration is not very efficient in the Balmer shocks surrounding \remnosp.  something about progenitor companion...no words left.. }

     %  for the baseline model (mono-atomic ideal gas, $E=1.4\times10^{51}\, \text{erg}$, and $M_{\rm{ej}} = 1.4 M_{\odot}$)}

     % The velocities are uniform along the eastern rim, averaging 6200 $\pm$ 180 km s$^{-1}$, while the southwestern rim shows significant fluctuations ranging from 5100 km s$^{-1}$ to 6700 km s$^{-1}$.  

\end{abstract}
\vspace{0.4in}
\vspace{0.4in}

\section{Introduction}
\label{sec:intro}

Supernova remnants (SNRs) are rich astrophysical laboratories for studying supernovae (SNe) and their effects on the interstellar medium (ISM). They arise from the interaction of the debris of an exploding star (supernova) with the ambient medium, causing powerful shocks and a recycling of the stellar material otherwise locked in stars. They are a direct probe of astrophysical shocks and the chemical enrichment of the ISM, and their fast ISM shocks are plausible sites for the acceleration of galactic cosmic rays \citep[see review by][]{bykov+18}. Remnants also allow us to study the explosion physics and the nature of the supernova progenitor. Having the means to study these processes becomes particularly necessary for Type Ia supernovae (SNe Ia), which are vital to modern cosmology as tools for measuring extra-galactic distances. SNe Ia have led to the discovery of cosmic acceleration, changing our very understanding of the expanding universe \citep{riess1998,perlmutter1999}. The potential plethora of information available from remnants makes them prime targets for studying several questions fundamental to modern astrophysics. \par 

SNRs in the Milky Way and nearby galaxies are snapshots of their SNe anywhere from a few hundred to several thousand years after the explosion. Structurally, they are bound by two shock features, namely the forward shock (FS) or blast wave expanding into the ISM, and a reverse shock (RS) that is a reaction to the expanding ejecta's deceleration by the ISM. The two shocks, along with the contact discontinuity (CD) that separates the ejecta from the shocked ambient medium, serve as the three main fluid discontinuities, whose locations and motions with respect to the explosion site can be used to determine a remnant's dynamical state. Accurate kinematic information obtained from high-quality data allows us to infer the properties of a SNR, thereby connecting an individual remnant to its originating SN. Our motivation for this paper follows on these lines, where we use excellent \HST imaging data to measure the FS velocity and angular radius of SNR \remnant and use it to measure the properties of its parent explosion and the local environment.  \par 

    \begin{figure*}[ht]
        \centering
        \hspace*{-0.2cm}
        \includegraphics[scale = 0.35]{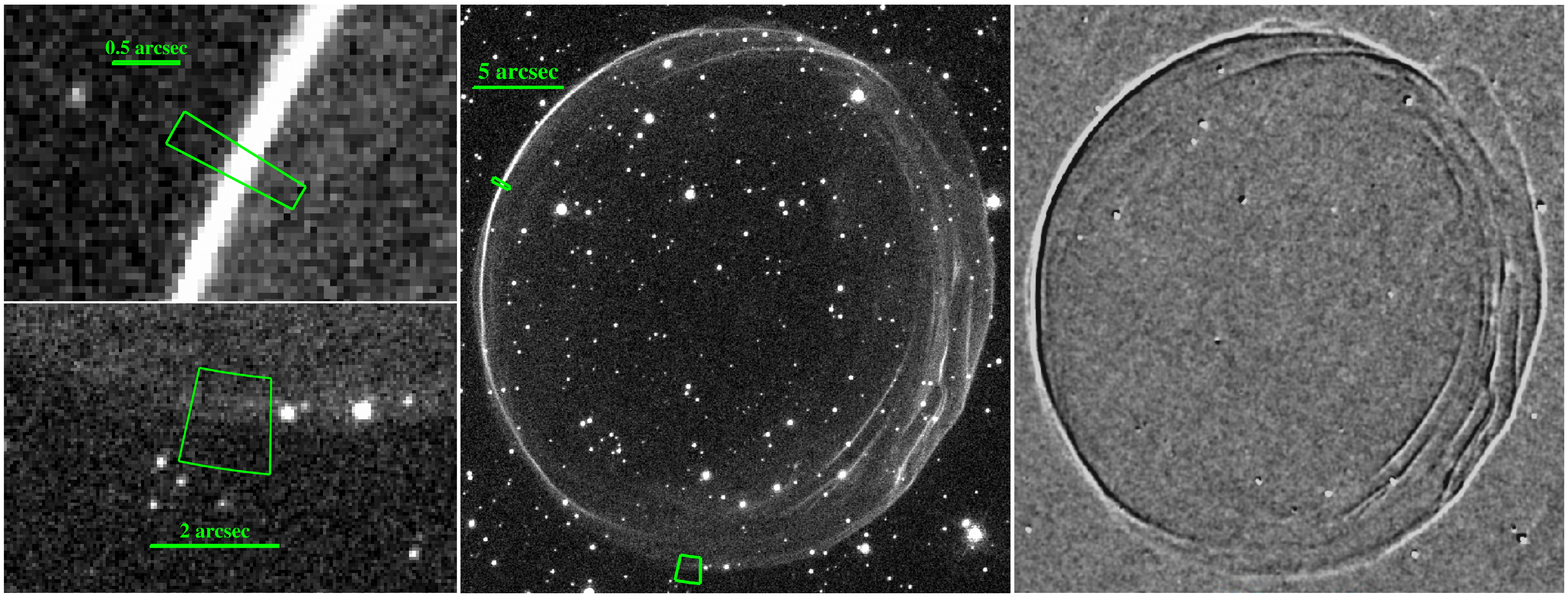}
        \caption{\textit{Center}: SNR \remnant \ha image showing the two types of apertures used for the proper motion analysis. \textit{Left}: Zoomed images showing a $1^{\circ}$-wide aperture (top) used for bright portions of the rim and a $5^{\circ}$-wide aperture (bottom) used for faint regions. \textit{Right}: Difference image of 0509$-$67.5 from observations separated by 10 years, clearly showing the remnant’s expansion (white shows regions where intensity increased, black shows where it decreased). }
    \label{fig:0509}
    
    \end{figure*} 

SNR \remnant is a young remnant located in the Large Magellanic Cloud (LMC), which exhibits remarkable symmetry in its morphology. It is one of only two SNRs that have been spectroscopically confirmed to be of SNe Ia origin (the other is the Tycho SNR, see \citealt{2008Natur.456..617K}), with light echo spectroscopy pointing to the highly energetic, over-luminous, 1991T-like SN Ia subtype \citep{rest2005,rest2008}. This subtype designation has been further strengthened from hydrodynamic modeling and X-ray spectral analysis of the remnant by \citet{badenes2008}, who estimated an ejecta mass of $M_{\rm ej} = 1.37 M_{\odot}$ and explosion energy of $E_{\rm kin} = 1.4 \times 10^{51}$ erg. Using proper motions (PMs) of the FS from two \HST images of \remnant observed about 1 year apart, \cite{hovey2015} (hereafter HHE15) employed 1D hydrodynamic simulations to constrain the age of \remnant to a value of $310\pm40$ yr for an assumed explosion energy of $1.4\times10^{51}$ erg. \par

Optical spectra of \remnant show strong Balmer emission, with no forbidden emission associated with radiative cooling, making it a Balmer-dominated (BD) remnant \citep{tuohy1982}. The \ha line, comprising a broad and narrow component, arises as both neutral and ionized hydrogen atoms from the ambient material enter the FS. BD shocks are ideal for studying the partition of shock energy into thermal populations of electrons and ions and relativistic particles, i.e., cosmic rays (CRs). Existing BD shock models \citep[]{vanadelsberg2008,morlino2013c} allow one to relate the measured broad \ha line width to the shock speed under different assumptions about the degree of electron to ion temperature equilibrium and the extent of CR acceleration. While these models have historically been used to estimate shock speeds from measured broad H$\alpha$ line widths, it has now become possible to constrain the efficiency of CR acceleration with accurate measurements of the shock speed and H$\alpha$ line width as shown by \citet{hovey+18}.    \par

While the H$\alpha$ observations of BD remnants are a direct probe of the FS location, there has been no direct observable for the RS. Its location has been usually inferred from the morphology of shock-heated ejecta, generally observed in the X-rays \citep{Hughes2003,warren2004, warren2005}, with instruments like \textit{Chandra} and \textit{XMM-Newton}. The location of the RS is calculated by fitting the radial Fe K$\alpha$ emission profile with spherically symmetric shell models and associating the inner radius of the shell with the RS. These fits, in addition to being simple geometric fits, are subject to large errors due to the limited knowledge of the precise charge state of Fe being modeled, given the CCD-type spectral resolution of the current X-ray instruments.   This concern, on the other hand, is significantly mitigated with optical spectroscopy of the RS heated ejecta, if one is able to identify ion species that result in optical emission from near the inner edge of the hot thermal plasma emitting X-rays. One such line is the [Fe XIV] $\lambda$5303 'coronal' line. \par 
A map of the [Fe XIV] coronal line was reported by \citet{seitenzahl+19} using integral field unit (IFU) spectroscopy with the Multi-Unit Spectroscopic Explorer (MUSE) instrument on the Very Large Telescope (VLT). The Fe$^{13+}$ charge state is intermediate between the low charge-state ($<$ Fe$^{2+}$) unshocked Fe ejecta in the interiors of SNe Ia remnants (for SN1006, see \citealt{wu1993faint}) and the high charge-state ($>$ Fe$^{16+}$) shocked material producing the bulk of the X-ray emission \citep[e.g.,][]{warren2004}. Since the Fe$^{13+}$ ion is short-lived in the RS zone, the emission from this line comes from a narrow range of radii (thin shell) where the temperature and ionization conditions are appropriate. A precise position of the RS relative to the coronal line location can be determined by modeling the coronal emission for an evolving SNR. Models from \citet{seitenzahl+19} indicate that the radius of the RS lies between 96 -- 97.5\% of the [Fe XIV] location. Thus, the coronal emission, corrected for this offset, is an accurate proxy for the RS position. \par

    \begin{figure*}[ht]
        \centering
        \hspace*{-0.2cm}
        \includegraphics[scale = 0.55]{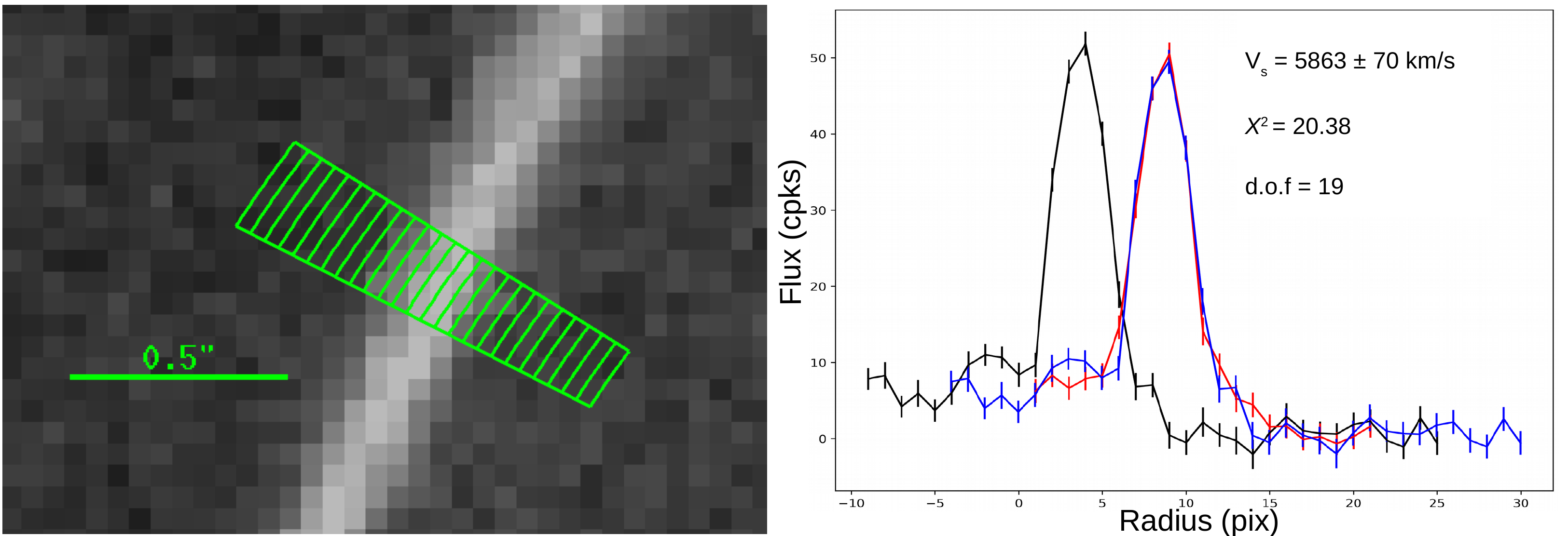}
        \caption{ \textit{Left}: Sample of a $1^{\circ}$-wide aperture used to extract a brightness profile. \textit{Right}: Brightness profiles of the extracted region; black shows the brightness profile from the 2006 observation, red from the 2016 observation, and blue shows the best-fit shifted profile. }
    \label{fig:bprofile}
    \end{figure*} 

In this paper, we use narrow-band H$\alpha$ \textit{HST} observations of \remnant taken $\sim$10 years apart to determine the  angular radius and PM expansion of the remnant's FS at numerous azimuthall positions around the rim.  We assume a distance to the LMC of $49.59$ from  \citet{Pietrzynski+2019}, and correct for the inclination of the LMC out of the plane of the sky using the relevant geometry from  \citet{vandermarel+2002}. With this $-0.32$ kpc correction, we end up with a distance of $49.27$ kpc (distance modulus of $18.463$) to SNR \remnosp. For the distance uncertainty, we use the thickness of the LMC reported by \citet{Choi+2018} of 1 kpc.
% Info from Pietrzynski+2019
% Distance modulus = 18.477 ± 0.004 (stat) ± 0.026 (sys)
% Distance = 49.59 ± 0.09 (stat) ± 0.54 (sys) kpc
% \sout {The current best distance estimate to the LMC of $50.12 \pm 0.46$ kpc} \sout{ (distance modulus $18.50 \pm 0.02$, see \citealt{alves04}) allows us to determine the kinematic properties of the FS to high} 
% \sout {precision. \citet{alves04} reports a LMC disk thickness of 1 kpc, which is the value we use for the distance uncertainty} \sout{ in our work. }

The \HST FS measurements are used to set constraints on the age, location of the dynamical explosion center, and the ambient-medium density via a Markov chain Monte Carlo (MCMC) analysis. Furthermore, using the results of the RS location from \citet{seitenzahl+19}, we determine the kinetic energy of the explosion. We consider the nonlinear effects of diffusive particle acceleration on the evolution of \remnant through the use of an effective $\gamma$ for the equation of state at the FS. We restrict our study to values of $\gamma{_\text{eff}} = 4/3$ and 5/3 corresponding to FS compression factors of 7 and 4, respectively.\par 

This paper is organized as follows: Section \ref{sec:datareduction} discusses the observations and data reduction of the \HST images of \remnosp. Section \ref{sec:pm} describes the methods used in calculating the PM of the FS. The model for the remnant's evolution, the model parameters, and the MCMC setup are presented in \S \ref{sec:analysis} and the MCMC results in \S \ref{sec:results}. In \S \ref{sec:discussion}, we consider the implications of our results, and finally in \S \ref{sec:conclusions}, we summarize and discuss future work. We use 1-$\sigma$ uncertainties for plots and quoted values in tables and the text unless explicitly noted otherwise. All \HST images were generated with a pixel scale of \scale. For the LMC distance estimate of 49.27 kpc, $1^{\prime\prime}$ (1 \HST pixel) corresponds to  0.239 (0.012) pc.\par

\section{Observations and Data Reduction}
\label{sec:datareduction}

For our analysis, we use two narrow-band \ha images of \remnosp, which were observed with the F658N filter on the Advanced Camera for Surveys (ACS). The first observation (hereafter epoch-1) was taken on 2006 October 28 for an exposure time of 4620 s under HST program 11015 (PI: Hughes). The second observation (hereafter epoch-2) was taken on 2016 November 13 with an exposure of 5331 s under HST program 14733 (PI: Hovey). SNR \remnosp, approximately $30''$ in diameter, was centered on CCD-2 of the WFC channel, which has a field of view of $202''$ x $202''$. To minimize the systematic errors in our results arising from unmatched geometric distortions in the plate scale and asymmetries in the point-spread function, the epoch-2 image was observed in the same orientation (roll-angle) as epoch-1. \par 

We analyze the \HST  pipeline pre-processed individual exposures (file suffix `flc’), that are calibrated and corrected for flat-fielding, bias, dark current, and charge transfer inefficiency (the latter corrected as described in \citealt{2010PASP..122.1035A}). We register all the images to a common WCS using the \texttt{tweakreg} task in the STScI  DrizzlePac package to (1) register an individual exposure to the Gaia (DR2) astrometric reference system \citep{gaia_2_2018, gaia2018}, and (2) register all the other exposures to this initial Gaia-registered exposure. For fine-tuning the object detection settings in \texttt{tweakreg}, we run the IRAF  \texttt{daofind} task to calculate the standard deviation of the sky background, and set the source detection threshold at 15.0 sigma above the background value. We run \texttt{tweakreg} with `rscale' geometry, which fits for an offset, a scale, and a rotation.  \par 

After registering the images, we use \texttt{astrodrizzle} which aligns and combines all the images in an output frame, after correcting for distortion, and identifying and removing cosmic rays. For sky subtraction, we use the default sigma-clipping routine incorporated in \texttt{astrodrizzle} and run it with a 4 sigma clipping factor using the median as the statistical parameter. We drizzle each input image separately onto an output frame, using a square kernel and an output pixel fraction of 0.8, which are then corrected for geometric distortion. The distortion-free output images are used to create a median image, which serves as the mask to remove cosmic rays. The relative registration uncertainties of the final drizzled epoch-1 and epoch-2 images have an RMS of $0.0065''$ and $0.0042''$ in the E-W and N–S axis directions,  respectively.\par

\section{Proper Motion Measurement}
\label{sec:pm}
\subsection{Methods}

The baseline for our PM measurement is 3669 days, during which time we see a significant expansion of the forward shock in the radial direction (see the right panel of figure~\ref{fig:0509}). There is little-to-no distortion in the shape of the shock front over this time; consequently, in the following analysis, we assume the local shock curvature does not change from one epoch to the other. Our analysis methodology broadly follows that of HHE15, where we extract 1D \ha brightness profiles from apertures defined on the shock fronts for both epochs. We differ from HHE15, however, in our use of ellipses to model the shape of the shock front, which traces its curvature  more accurately than the rectangular boxes used in HHE15. We define 36 elliptical apertures, each with an angular width of $10^\circ$, spanning the entire rim, matching the shape of the shock locally (aligned by eye). For the bright filaments, we further divide each elliptical region into 10 sub-regions, creating $1^\circ$-wide apertures. For the fainter portions of the rim in the south and northwest, we use $5^\circ$-wide apertures, in order to increase the signal-to-noise (S/N) in the extracted brightness profiles. Examples of apertures for both bright and faint emission regions are shown in figure~\ref{fig:0509} (left and middle panels). \par 

Before extracting the profiles, we ensure that the regions of extraction are free of stellar contamination. We use the Gaia and USNO catalogs to identify the stars in the field that are located on or around the shock. We discard all regions with stars co-located on the shock front in either epoch. Stars that fall within an extraction region, but not directly on the shock-front, are masked. This process leaves us with 231 viable regions, for which we determine the local background flux for each epoch's image separately. This is done by sampling the source-free background region upstream of the shock in each aperture, and fitting the resulting pixel intensity distribution to a Gaussian. For each region, we use the mean and the standard deviation of the Gaussian as the background value and its uncertainty respectively. Typical background values are $3.7 \times 10^{-3}$  s$^{-1}$ and $6.9 \times 10^{-3}$  s$^{-1}$ for epoch-1 and epoch-2 respectively. \par 

We extract the fluxes and inverse-variance weights for all pixels that fall within a given aperture, binning them by their radial pixel positions (see figure~\ref{fig:bprofile}, left panel). We determine the weighted average of the flux within each bin and then subtract the corresponding background value. Figure~\ref{fig:bprofile} (right panel) shows an example of the background-subtracted \ha brightness profile for a bright filament in the eastern rim, where the black and red curves correspond to the epoch-1 and epoch-2 profiles respectively. Due to the superior instrument performance and S/N for the earlier observation, we treat the epoch-1 profiles as our model ($M_i$, with uncertainty $\sigma_{M_{i}}$), and epoch-2 profiles as the data ($D_i$, with uncertainty $\sigma_{D_{i}}$), where $i$ is the index corresponding to each radial bin. We estimate the best fit integer shift value and intensity scale factor ($F$), for the model profile to match the data profile using the chi-square figure-of-merit function, defined as

\begin{equation}
 \chi_{k}^{2}=\sum_{i=1}^{N} \frac{\left(D_{i}-M_{i+k} F_k\right)^{2}}{W_{ik}}\ ,   
\end{equation}

where $W_{ik} = \sigma_{D_{i}}^{2}+\langle F\rangle^{2} \sigma_{M_{i+k}}^{2}$, 
\(\langle F\rangle=\sum D / \sum M\) (the average brightness scale factor), and

\begin{equation}
F_k=\cfrac{\sum_{i=1}^{N} M_{i+k} D_{k}/W_{ik}} 
          {\sum_{i=1}^{N} M_{i+k}^{2}/W_{ik}} \ . 
\end{equation}

We interpolate on the chi-square values for each integer shift ($k$) of the model profile using a cubic spline, and minimize the interpolated function to determine the best-fit shift value ($S_b$). We explored an alternative approach of interpolating on the brightness profiles themselves, but find that cubic-spline interpolation is unable to accurately follow the profile shapes (some of which vary strongly from bin to bin). The blue curve in figure~\ref{fig:bprofile} (right panel) shows the shifted and scaled epoch-1 profiles matched with the epoch-2 profile in red. We calculate the $1\sigma$ uncertainties on $S_b$ using the $\Delta \chi^2 = 1.0$ criterion. We also estimate the systematic uncertainties from our fitting method by fitting the $\chi_{k}^2$ values of the 3 integer shift positions closest to $S_b$ with a parabola and take as the systematic uncertainty the difference in the minimum of the parabola fit and the interpolated function. In Table \ref{tab:da_results} (column 4), we give the $S_b$ values from  selected apertures as the angular shift (outward-going radial shifts are quoted as positive). We also quote a total uncertainty value equal to the root-sum-square of the statistical and systematic uncertainty. 
% We also quote the shift values as a FS velocity  ($V_{FS}$)} in column 5 using the LMC distance of   49.27} kpc.
The minimum $\chi_{k}^2$ and the degrees of freedom of each fit are given in columns 5 and 6 respectively. \par 

 Next, we measure the position of the FS at each aperture. To do this, we obtain from the epoch-2 brightness profiles, the X and Y positions of the peak surface brightness, and the position where the flux of the outer edge of the profile falls to half the peak value. We temporarily convert these positions to radii ($R_{\rm max}$ and $R_{1/2}$ respectively) using the {\it geometric center} of $\alpha_{\rm{geo}}= 5^{\rm h}09^{\rm m}31^{\rm s}.16$ and $\delta_{\rm{geo}} =  -67^{\circ}31^{\prime}17.1^{\prime\prime}$, which we determine by fitting the X and Y locations of the peak (or the half peak) with an ellipse. Assuming the projection of a uniformly-emitting, spherical, geometrically-thin shell for the \ha shock profile, the FS radius is given by

% The radius of the FS determined from the epoch-2 image assuming  the projection of a} uniformly-emitting, spherical, geometrically-thin shell for the \ha shock profile. Geometry gives the FS radius as

\begin{equation}
    {R_{\rm FS} = \sqrt{\frac{4R_{1/2}^2 - R_{\rm max}^2 }{3}} - \frac{S_b}{2}}\ .
\label{eqn:rfs}
\end{equation}
% where $R_{\rm max}$ is the radial position of the peak surface brightness and $R_{1/2}$ is the position where the flux of the outer edge of the profile falls to half the peak value.

In the above equation, we correct the radii values to correspond to the midpoint in time of the two epochs by subtracting half the best-fit shift values from the epoch-2 FS radii (since the measured FS velocity is the average velocity over the time between the two epochs). This correction ensures that both the FS position and velocities are quoted at the same time in the remnant's evolution (specifically November 2011).  This is necessary for our subsequent analysis, where we precisely determine the explosion properties of \remnosp.  Finally, using the $R_{\rm FS}$ values determined from equation \ref{eqn:rfs} and the position angles (hereafter PA, from the geometric center) to the center of each aperture (Table \ref{tab:da_results} column 7), we calculate the FS location. The RA and DEC values of the FS location at each of the selected apertures are shown in columns 1 and 2. Later, we will use them as starting points to determine the center of the explosion. The uncertainties in the radii (see column 3) are generously quoted as the root-sum-square combination of $R_{1/2} - R_{\rm max}$ and the uncertainty in the best-fit shift values. Note that the location of the FS with respect to the locations of the $R_{\rm max}$ and $R_{1/2}$ is independent of the choice of center. We verify this by varying the center around the geometric center by several arcseconds and find that the variation in FS location is at least an order of magnitude lower than the radius uncertainty, and hence can be ignored. \par
% The radii values and their corresponding uncertainties are shown in Table \ref{tab:da_results}. 

\begin{table*}[t]
\centering
\setlength{\extrarowheight}{6pt} 
\setlength\tabcolsep{9pt}
\begin{threeparttable}
\caption{Kinematic measurements of the FS position and velocity (quoted as a shift) of SNR \remnant }
\begin{tabular}{c c c c c c c c}
\toprule
 R.A.(FS) & Dec.(FS) & $\sigma_{\rm FS-rad}$ & Rim PM$^{\left(1\right)}$   &  ${\chi^2}$ & d.o.f & PA$^{\left(2\right)}$ & $\rho^{\left(3\right)}$ \\ 
 (J2000)& (J2000) & $('')$ & (pix) &  &  & (deg) & ($\times 10^{-25}$ g cm$^{-3}$)\\ 
\midrule
77.382408 & -67.517219 &  0.10 & 5.62 $\pm$ 0.19 & 16.7 & 19 & 13.19 &  5.00\\
77.386354 & -67.518036 &  0.13 & 5.31 $\pm$ 0.11 & 20.0 & 19 & 36.36 &  6.07\\
77.389596 & -67.519711 &  0.07 & 4.96 $\pm$ 0.06 & 20.4 & 19 & 65.36 &  7.46\\
77.390475 & -67.521931 &  0.23 & 5.15 $\pm$ 0.17 & 28.9 & 20 & 97.12 &  7.16\\
77.388550 & -67.524086 &  0.28 & 4.77 $\pm$ 0.32 & 27.8 & 19 & 128.63 &  6.04\\
77.381654 & -67.525833 &  0.14 & 6.62 $\pm$ 0.38 & 26.2 & 20 & 171.03 &  3.85\\
77.374092 & -67.525122 &  0.31 & 5.65 $\pm$ 0.22 & 14.4 & 19 & 210.67 &  4.89\\
77.371538 & -67.523936 &  0.10 & 5.06 $\pm$ 0.24 & 25.1 & 20 & 231.59 &  7.66\\
77.369283 & -67.522006 &  0.19 & 5.64 $\pm$ 0.3 & 12.1 & 19 & 261.76 &  7.03\\
77.373929 & -67.517717 &  0.32 & 5.63 $\pm$ 0.34 & 25.2 & 19 & 328.64 &  4.80\\

\bottomrule
\end{tabular}
\label{tab:da_results}
\begin{tablenotes}\footnotesize
\item  Note: This table is just a short descriptive version of full table provided as an online data component.  $(1)$ Rim proper motion in the shock-normal direction, 1 HST pixel = 50 mas. $(2)$ Position Angle measured at the center of each sector assuming the geometric center. $(3)$ The density values correspond to the MCMC analysis presented in \S \ref{sec:fitted_center}.
\end{tablenotes}
\end{threeparttable}
\end{table*}

    \begin{figure*}[t]
        \centering
        \hspace*{-0.8cm}
        \includegraphics[scale = 0.81]{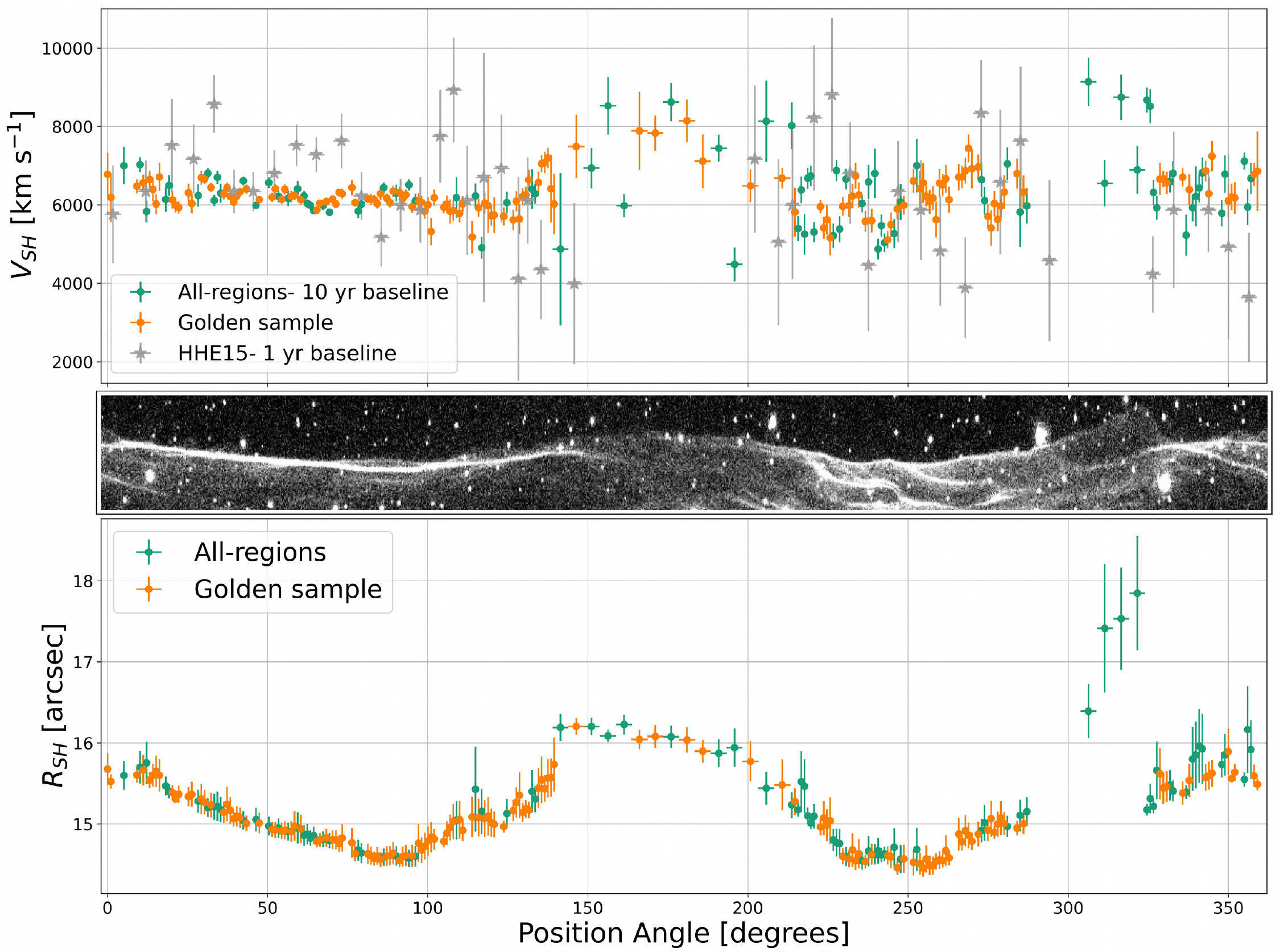}
        \caption{Velocity and radius of the FS of SNR \remnant as a function of position angle around the rim. The shock velocities and radii for all the 231 regions are shown in green in the top and bottom panel respectively, while the golden sample (see Sec:\ref{sec:GS}) is over-plotted in orange. In the top panel, we also show the shock-velocity results from \citet{hovey2015} (plotted in gray), measured over a 1-yr baseline. The longer baseline by almost a factor of 10 for our data allows us to measure the proper motion  of the rim to higher precision, while also showing the variability over small spatial scales. The middle panel shows an image of SNR \remnosp's shock front region in polar coordinates (radius versus azimuthal position angle) aligned with the other panels along the position angle axis.  }
        \label{fig:velradpa}
    \end{figure*} 

\subsection{Determining a Golden Sample}
\label{sec:GS}

Before interpreting our measurements of the radii and the PMs in the context of SNR evolution, we perform a series of quality cuts on all 231 apertures to create a golden sample for subsequent analyses. The first cut is based on the velocities of neighboring (in terms of PA) apertures through a moving-average calculation. We determine the weighted mean and associated standard uncertainty of the velocities from four regions adjacent to a given region (two ahead and two behind). We include the middle region in the golden sample if its velocity is within $\pm 2 \sigma$ of the neighboring ones.  This quality cut rests on the argument that regions that are physically close to one another should exhibit similar shock velocities given the smooth structure of the FS. We experimented with the number of adjacent regions to use for this estimate: two was optimal to avoid including too much fluctuation in the velocities versus washing out small-scale velocity features.  \par 

The second quality cut is based on the profile fits and employs the $\chi^2$ probability-to-exceed (PTE) the null hypothesis (i.e., that the model describes the data). The $\chi^2$ PTE rejection ensures only those regions where the shock profiles have retained their shape over the course of the remnant's expansion are being used, justifying the assumption we made at the beginning of our analysis. Here we require that the PTE be $<$95.4\% (also $2 \sigma$). \par 

The third quality cut is also based on the profile fits.  Here we compute the mean and RMS of the fitted brightness scale factor ($F_k$ at the best $k$ value) for all apertures and make a $\pm 2 \sigma$ outlier cut ($2 \sigma$ range is $1.01\pm 0.18$ ). This ensures that we only have regions where the brightness of the shell has not changed significantly over the time baseline of our measurements. \par 

In contrast to the three preceding cuts, the final cut is based on the shock radii. We find that for certain regions, the outermost shock front is accompanied by a closely spaced fainter shock. This introduces an uncertainty in the value of $R_{1/2}$, thereby affecting the value of $R_{FS}$. Therefore we remove regions with shock thickness $2 \sigma$ greater than the average shock thickness of all apertures. \par 

Applying all the quality cuts eliminates 87 regions from the original set of 231, leaving 144 regions that comprise the golden sample. We note here that there is a significant overlap in the regions identified as outliers in the first three cuts, and there are no discrepancies in our list of outliers if any of the above cuts are performed in a different order. \par

Plots of the shock velocity and radii as a function of PA around the rim are given in figure~\ref{fig:velradpa}.  A polar plot of the outer rim of the \HST\ image is shown in the middle panel for  ease of comparison to figure~\ref{fig:0509}. The measured quantities refer to the outermost extent of the remnant; interior filaments (such as those around azimuths of 250$^\circ$) are not included in this study.  This figure clearly shows the lower scatter of the golden sample compared to the original full sample. 
\par

\section{Analysis Methods}
\label{sec:analysis}
We analyze our kinematic measurements in the context of hydrodynamic simulations of SNR evolution into uniform density ambient media to set constraints on the dynamical parameters of \remnosp. The hydro code determines the radial profiles of gas properties (e.g., density, temperature, velocity) as a function of time, from which the time evolution of the shock radii and velocities are extracted. These functions are reduced to dimensionless form by scaling with appropriate characteristic values for radius, velocity, and time. The use of scaling laws allows for the hydro evolution to be efficiently implemented in an MCMC framework; here we employ the Metropolis-Hastings algorithm to sample through the probability space. Given a set of model parameters, the MCMC code evaluates the likelihood of the modeled radius and velocity given the measured values to generate the posterior probability distribution. These posteriors yield the optimized, maximum-likelihood values for the dynamical parameters along with their confidence intervals. For this work, we use the \texttt{`emcee'} python package developed by \citet{emcee}. In the following, we describe in more detail the hydro model, the radius and velocity scaling laws, and the set-up of the MCMC code. We use as our primary data set the radii and velocity measurements and respective uncertainties from the golden sample (\S \ref{sec:GS}) that come from 144 independent azimuthal apertures around the rim. 

% Our analysis introduces global parameters that characterize the explosion and have the same value for all PA, and a local parameter (ambient medium density) that is allowed to vary independently for each azimuth.

\subsection{Model}
 Following HHE15, we employ the hydro-code Darla, which solves the Euler equations for a 1D Lagrangian spherically-symmetric mesh. We use an exponential ejecta density profile, which provides a better match than power law profiles to the density profiles of various simulations of different explosion models \citep{dwarkadas1998}. The ejecta density and velocity profiles depend on the ejecta mass ($M_{\rm{ej}}$) and ejecta kinetic energy ($E_{51}$, which we write in units of $10^{51}$ erg). Assuming that the ejecta are expanding into a uniform ambient medium ($\rho$), we describe the remnant evolution using a similarity solution \citep{1973MNRAS.161...47G, mckee1995explosions}, given by 

\begin{equation}
\begin{aligned}
\label{eqn:chdim}
R &= R_{\rm ch}\  r[t] \  \\
V &= V_{\rm ch}\  v[t] \ \\
T &= \dfrac{R_{\rm ch}}{V_{\rm ch}}\  t \ ,
\end{aligned}
\end{equation}
where $R$, $V$ and $T$ are the remnant's radius, velocity and age, written in terms of the scaled (dimensionless) functions $r[t]$, $v[t]$, and $t$, respectively. The characteristic (scaling) values in these equations are given by the following 
\begin{equation}
\begin{aligned}
\label{eqn:ch}
   R_{\rm ch} &= \bigg (\frac{3M_{\rm ej}}{4 \pi \rho}\bigg )^{1/3}  \\
  V_{\rm ch}  &= \bigg (\frac{2 E_{51}}{M_{\rm ej}}\bigg )^{1/2} \\
T_{\rm ch} &= \frac{R_{\rm ch}}{V_{\rm ch}} = M_{\rm ej} ^{5/6} (4\pi \rho/3)^{-1/3} (2 E_{51})^{-1/2} \\
\end{aligned}
\end{equation}

 For given values of  $M_{\rm{ej}}$, $E_{51}$ and $\rho$, Darla calculates the positions and velocity of the FS, reverse shock (RS), and the contact discontinuity (CD) for each time step of the simulation. We convert each of these quantities to their dimensionless counterparts using the characteristic values from equation \ref{eqn:ch}. 

\begin{figure}
            \centering
            \includegraphics[scale=.42]{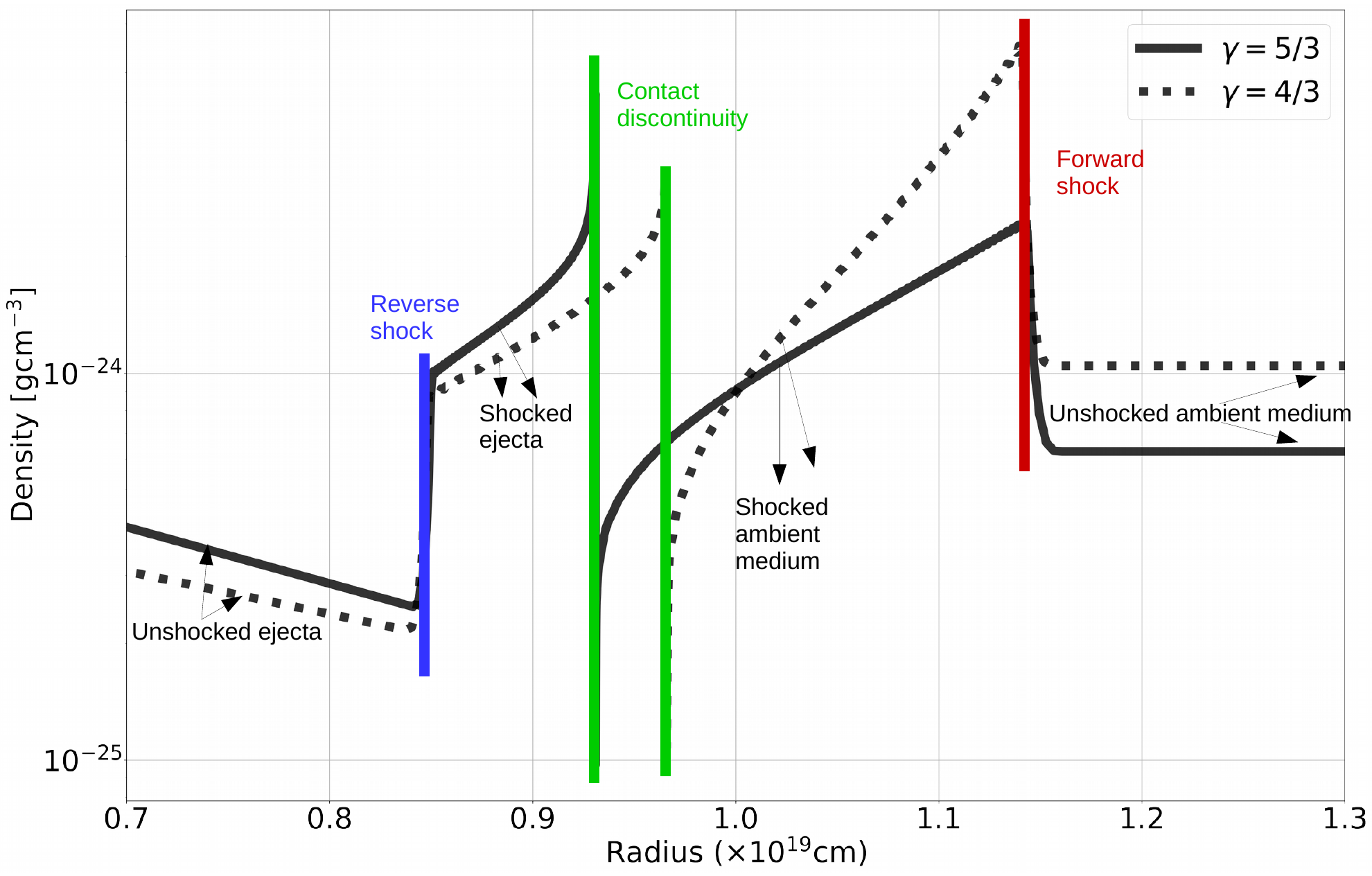}
            \caption{Radial density profiles for $\gamma_{\rm eff} = 5/3$ (solid curve) and $\gamma_{\rm eff} = 4/3$ (dotted curve) reflecting the compression factors for the two cases. The profiles are shown for the age, density and energy values corresponding to rows 1 and 3 of Table \ref{tab:mcmc_energy_results} for a fixed mass of $M_{\rm{ej}} = 1.4 M_{\odot}$. }
            \label{fig:shock}
        \end{figure}

To approximate the effects of diffusive-shock acceleration (DSA) in the hydro simulations, we introduce an effective adiabatic index ($\gamma_{\rm eff}$) for the upstream ambient material, which artificially modifies the shock compression factor at the FS.  The strong-shock compression factor is given by $(\gamma_{\rm eff} +1)/(\gamma_{\rm eff} -1)$. In this paper, we consider two cases $\gamma_{\rm eff} = 4/3$ and $5/3$ (yielding compression factors of 7 and 4), which correspond to the equations of state for relativistic and non-relativistic ideal monatomic gases. Figure~\ref{fig:shock} shows an illustration of the compression factors achieved for these values of $\gamma_{\rm eff}$.  The consequences are evident in the different  density jumps at the FS (red vertical bar) and the separation between the FS position and the corresponding locations for the CD \footnote{Although the location of the contact discontinuity is an important marker for the SNR dynamics, observations of it are difficult because it is dynamically unstable, see \citealt{1973MNRAS.161...47G}.} (green bars) for the two cases. Note that we do not modify the equation of state of the ejecta, so the location of the RS is not affected by our approximate inclusion of DSA in the simulations. Although there have been theoretical studies exploring the effects of DSA at RSs \citep{Ellison2004,Telezhinsky2012}, there is no direct evidence from observations of remnants \citep[with the possible exception of Cas A see, e.g.,][]{sato+2018}.  \par

 The outputs of the hydro code consist of tabulated values for radius and velocity as a function of time, which we convert to dimensionless forms, e.g. $r[t]$ and $v[t]$. We find that both $r[t]$ and $v[t]$ show numerical noise of order $\lesssim 10^{-3}$. To eliminate these noise fluctuations, we fit $r[t]$ with a 16th-order and $v[t]$ with a 15th-order polynomial, which faithfully characterize the evolution of shock radius and velocity with time. These polynomial functions allow us to easily implement the model (equations \ref{eqn:chdim} and \ref{eqn:ch}) in our MCMC analysis.

% We remove small-scale numerical noise in the $r[t]$ and $v[t]$ functions output from the hydro code by fitting them with  16th- and 15th-order polynomials respectively.

% The polynomial fits result in minimal residuals compared to the original functions, while faithfully characterizing the evolution of shock radius and velocity with time. 

% This process was necessary because tiny noise variations in the model radius and velocity values were noticeable in the fits to the {\it HST} data,  given its high precision. \par 

\subsection{MCMC set-up and inference: comparing model with data}

 Our 1-D hydro simulations assume a spherically symmetric explosion, and thereby produce a spherical SNR. However, SNR \remnant has a clearly elliptical shape (with major/minor axis approximately N-S/E-W). Consequently, there needs to an asymmetry in some quantity describing the SNR that varies with PA. In our analysis moving forward, we consider the variables allowed to vary with PA as {\it local} parameters, while the rest (unchanging with PA) are {\it global} parameters. Our baseline model involves the following global parameters describing \remnosp: age ($T$), distance ($D$), location of the explosion ($\alpha_{\rm{cen}}$ and $\delta_{\rm{cen}}$), ejecta kinetic energy ($E_{51}$), and ejecta mass ($M_{\rm{ej}}$). We allow the ambient density ($\rho_i$) to vary with PA, and determined independently for each azimuth (indexed by the label $i$). This makes the ambient density as a local parameter, accounting for the varying size of the elliptical remnant. The correlation of the brightness of the rim with the radius (E-W: brighter rim/smaller radii, N-S: fainter rim/larger radii, see fig 3 bottom panel) argues strongly for a density variation around the rim. However, we also run a model where we fix the density and allow the explosion energy to be the local parameter that varies around the rim. As we will show, this energy-varying case is statistically as good a fit as the density-varying one. We will discuss the results and implications of this in \S \ref{sec:densities}.

% in order to assess the systematic uncertainty in the explosion center. }

%  Explain how you are planning to account for the clearly elliptical shape of the remnant. Spherical explosion, spherical remnant. Explain that the solution we are producing the remnant. There needs to be an asymmetry of some quantity desctipbing the SNR that varies with PA. The variable that varies with PA , we consider local, while the others are global. Our}Our baseline model involves the following {\it global} input parameters describing \remnosp: age ($T$), distance ($D$), location of the explosion ($\alpha_{\rm{cen}}$ and $\delta_{\rm{cen}}$), ejecta kinetic energy ($E_{51}$), and ejecta mass ($M_{\rm{ej}}$).
%  In the baseline model, we will allow the density to vary with PA, which will account for the varying size of the elliptical remnant. The correlation of the brightness of the rim with the radius (EW: bright/smaller radii, NS:faint rim/larger radii, see fig 3 bottom panel) argues strongly for a density variation around the rim. However, we also run a model where we fix the density and allow the explosion energy to vary around the rim. We will discuss it in dynamical center section... in order to assess to give a systematic uncertainty. }

% We refer to these quantities as global because the same value of each is used for all position angles.  We include the ambient medium density as a local parameter ($\rho_i$) to be determined independently for each azimuth (indexed by the label $i$). 
The Gaussian figure-of-merit likelihood function ($\Lagr$) to be maximized is

\begin{equation*}
\log(\Lagr) = \log(\Lagr_{\dot{\theta}}\Lagr_{\theta}) =
-\sum_{i=1}^{n}\left[\log(2\pi\sigma_{\theta_i}^2) + \log(2\pi\sigma_{\dot{\theta}_i}^2)\right]  \\
\end{equation*}
\begin{equation}
\label{eqn:loglike}
-\sum_{i=1}^{n}\left[ \frac{\left(\theta_i -R_i/D\right)^2}{\sigma_{\theta_i}^2} + \frac{\left(\dot{\theta}_i-V_i/D\right)^2}{\sigma_{\dot{\theta}_i}^2}\right]\\
\end{equation}
where the variables $R_i$ and $V_i$ depend on the global and local parameters through equations \ref{eqn:chdim} and \ref{eqn:ch}. 

The quantities  $\theta$ and  $\dot \theta$  in the above equation are the radius and velocity values from each azimuthal aperture in angular units, thus allowing us to include the remnant's distance explicitly as a fit parameter. Furthermore, we determine the individual radius positions $\theta_i$ using
\begin{equation}
\theta_i = \big [ ( {\rm X}_{i} - {\rm X_{cen}} ) ^2 
   + ( {\rm Y}_{i} - {\rm Y_{cen}} ) ^2  \big ]^{1/2} \ ,
   \end{equation}
where ${\rm X}_{i}$ and ${\rm Y}_{i}$ are the image coordinates of the shock front  (equivalent to columns 1 and 2 of Table \ref{tab:da_results}) in azimuthal bin $i$, and ${\rm X_{cen}}$ and ${\rm Y_{cen}}$ are the image coordinates of the remnant's center. This refinement allows us to determine the remnant's center ($\alpha$ and $\delta$) based on the measured kinematics under the best-fit hydro model. \par 

It is not possible to constrain all six input global model parameters using only measurements of the FS. What the FS radii and velocity data can constrain are $T$, $\alpha_{\rm{cen}}$, $\delta_{\rm{cen}}$, and all $\rho_i$ values associated with each azimuth. We include the distance $D$, even though it is very well constrained, because its value correlates strongly with the age and density values.
\par

We set flat and wide priors on the age of $300 \pm 150$ yr, and a strong Gaussian prior on the distance with mean of $\mu =   49.27$ kpc and standard deviation of $\sigma = 1$ kpc (the thickness of the LMC disk, \citet{Choi+2018}). We cast the densities in log units, and set a wide flat prior: $-30<\log(\rho_i/{\rm{g}}\,{\rm{cm}}^{-3})<-20$. For the initial center location values, ${\rm X_{cen}}$ and ${\rm Y_{cen}}$, we use the geometric center. We set a flat prior on the position with a width of $\pm 10^{\prime\prime}$ in each coordinate.  This is a very generous prior corresponding to 2/3 of the average radius of \remnosp.

%  determined by fitting the FS radii with an ellipse, which yields the position $\alpha_{\rm{geo}}= 5^{\rm h}09^{\rm m}31^{\rm s}.16$ and $\delta_{\rm{geo}} =  -67^{\circ}31^{\prime}17.1^{\prime\prime}$.

\section{Results}
\label{sec:results}

Below, we discuss the results of three specific sets of MCMC runs using the golden sample of FS radius and velocity measurements. In the first subsection below, we measure the age and LMC distance (global parameters) and the ambient density within each of the 144 azimuthal sectors (local parameters). In the next subsection, in addition to these parameters, we also measure the R.A.\ ($\alpha_{\rm{dyn}}$) and Decl.\ ($\delta_{\rm{dyn}}$) positions of the remnant's dynamical center (global parameters). For both sections unless otherwise stated, we fix the ejected mass and explosion energy to $M_{\rm{ej}} = 1.4 M_{\odot}$ and $E_{51} = 1.4$.

In the third section, we include information on the position of the RS estimated from the [Fe XIV] $\lambda$5303 image obtained with ground-based \textit{VLT} MUSE data. This allows us to constrain an additional dynamical parameter, either the explosion energy or the ejected mass. We show results for two values of the effective adiabatic index ($\gamma_{\rm eff} = 4/3$ and $5/3$) in the upstream medium, which manifests as a change in the FS compression factor, thereby changing the dynamics of the FS and the location of the other fluid discontinuities. \par

\begin{deluxetable*}{c c c c c c c c c}[!b]
\tablecolumns{9}
\tablewidth{0pc} 
\tabletypesize{\scriptsize}
\tablecaption{MCMC posteriors for age, distance and dynamical center of SNR \remnant}
 \tablehead{
  \colhead{Center} & \colhead{$\gamma_{\rm eff}$} & \colhead{$T$ (yr)} 
    & \colhead{$D$ (kpc)}
    & \colhead{$\alpha_{\rm{dyn}} -\alpha_{\rm{geo}}\, ('')$ }
    & \colhead{$\delta_{\rm{dyn}} -\delta_{\rm{geo}}\, ('')$ } 
    & \colhead{${\chi^2}_{\rm{FSR}}$}  & \colhead{${\chi^2}_{\rm{FSV}}$} & \colhead{Total d.o.f}
  }
\startdata
\multirow{2}{*}{Fixed} & 5/3 &  $315.98\pm 1.7$ & $50.18\pm 1.0$ & \nodata & \nodata &  $36.3$ &  $245.1$ & $142$ \Bstrut{}\\
& 4/3 &  $316.42\pm 1.8$ &  $50.15\pm 1.0$ & \nodata & \nodata &  $34.5$ &  $246.2$ & $142$ \Bstrut{}\\

\hline
\bstrut{} \\
\multirow{2}{*}{Fitted} & 5/3 &  $315.52\pm 1.76$ &  $50.14\pm 0.95$ &  $-0.024\pm 0.049$ &  $0.012\pm 0.064$ &  $33.1$ &  $245.2$ & $140$ \Bstrut{}\\
\vspace{-0.5cm}
& 4/3 &  $316.05\pm 1.9$ &  $50.16\pm 0.99$ &  $-0.025\pm 0.048$ &  $0.015\pm 0.063$ &  $34.0$ &  $244.1$ & $140$ \Bstrut{}\\
\enddata
 \tablecomments{MCMC posteriors obtained for fixed values of $M_{\rm{ej}} = 1.4 M_{\odot}$ and $E_{51} = 1.4$, using the golden sample of 144 FS radius and velocity measurements. }
\label{tab:MCMC_tab}
\end{deluxetable*}
\subsection{MCMC with center fixed at the geometric center}
\label{sec:fixed_center}

%  \begin{deluxetable*}{c c c c c c c c c}[!b]
% \tablecolumns{9}
% \tablewidth{0pc} 
% \tabletypesize{\scriptsize}
% \tablecaption{MCMC posteriors for age, distance and dynamical center of SNR \remnant}
%  \tablehead{
%   \colhead{Center} & \colhead{$\gamma_{\rm eff}$} & \colhead{$T$ (yr)} 
%     & \colhead{$D$ (kpc)}
%     & \colhead{$\alpha_{\rm{dyn}} -\alpha_{\rm{geo}}\, ('')$ }
%     & \colhead{$\delta_{\rm{dyn}} -\delta_{\rm{geo}}\, ('')$ } 
%     & \colhead{${\chi^2}_{\rm{FSR}}$}  & \colhead{${\chi^2}_{\rm{FSV}}$} & \colhead{Total d.o.f}
%   }
% \startdata
% \multirow{2}{*}{Fixed} & 5/3 & $317.2\pm 1.7$ & $50.9\pm 1.0$ & \nodata & \nodata & $34.3$ & $246.6$ & 142 \Bstrut{}\\
% & 4/3 & $317.8\pm 1.7$ & $51.0\pm 1.0$ & \nodata & \nodata & $39.0$ & $241.8$ & 142 \Bstrut{}\\
% \hline
% \multirow{2}{*}{Fitted} & 5/3 & $316.7\pm 1.8$ & $50.9\pm 1.0$ & $-0.027\pm 0.050$ & $0.011 \pm 0.062$ & $33.4$ & $244.8$ & 140 \Bstrut{}\\
% & 4/3 & $317.6\pm 1.8$ & $51.0\pm 1.0$ & $-0.023\pm 0.049$ & $0.013 \pm 0.062$ & $34.0$ & $244.2$ & 140 \Bstrut{}\\
% \enddata
%  \tablecomments{MCMC posteriors obtained for fixed values of $M_{\rm{ej}} = 1.4 M_{\odot}$ and $E_{51} = 1.4$, using the golden sample of 144 FS radius and velocity measurements. }
% \label{tab:MCMC_tab}
% \end{deluxetable*}
% \subsection{MCMC with center fixed at the geometric center}
% \label{sec:fixed_center}

   \begin{figure*}[ht]
        \centering        
        \includegraphics[scale=0.45]{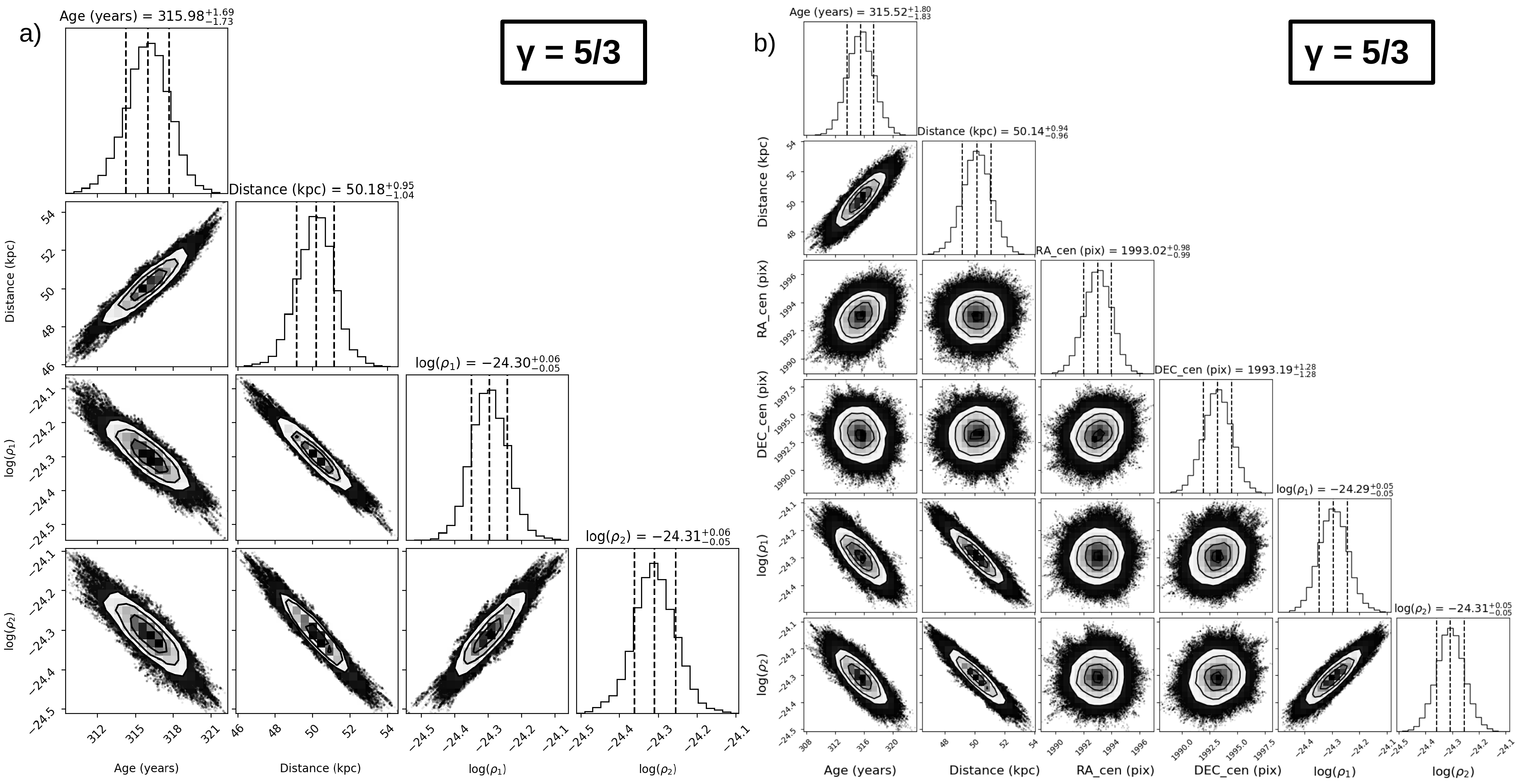}
        \caption{(a) Corner plot showing the posterior distributions of the age ($T$), distance ($D$), and two of the 144 densities ($\rho_i$) constrained by the MCMC analysis with \remnosp's center fixed. The densities correspond to position angles 0.34$^\circ$ and 10.32$^\circ$. (b) Corner plot for the case where we fit for the center, given by the two additional parameters $\rm{RA\_{cen}}$ and $\rm{DEC\_{cen}}$ (equivalent to $\alpha_{\rm{dyn}}$ and $\delta_{\rm{dyn}}$).}  
        \label{fig:mcmc_corner_fs}
    \end{figure*} 

 Table \ref{tab:MCMC_tab} (top panel) summarizes our results for this set of MCMC runs where the center is kept fixed at the geometric center and we vary the age, distance, and 144  ambient medium density values. The remnant's age turns out to be  $317.2 \pm 1.7$ yr and the distance is $50.9 \pm 1.0$ kpc. The age values are consistent with HHE15's value of $310\pm40$ yr, but with much smaller uncertainties due to our larger number of azimuthal data points and more accurate velocity measurements made possible by the $\sim$10 times longer time baseline between observing epochs. In figure~\ref{fig:mcmc_corner_fs}(a), we show the posterior distributions and the covariances between $T,D$ and $\rho_i$ for two azimuthal regions (at PA 0.34$^\circ$ and 10.32$^\circ$). The posteriors for $T$ and $D$ are approximately normally distributed, while the $\rho_i$ distributions are approximately log-normal. The remnant's age is strongly correlated with distance, while the $\rho_i$ are strongly anti-correlated with $D$, which introduces large correlated uncertainties in their estimated values.  If we fix the distance, we find that the densities are only weakly correlated with age.  \par
 
     \begin{figure*}[t]
        \centering
        \includegraphics[scale=0.63]{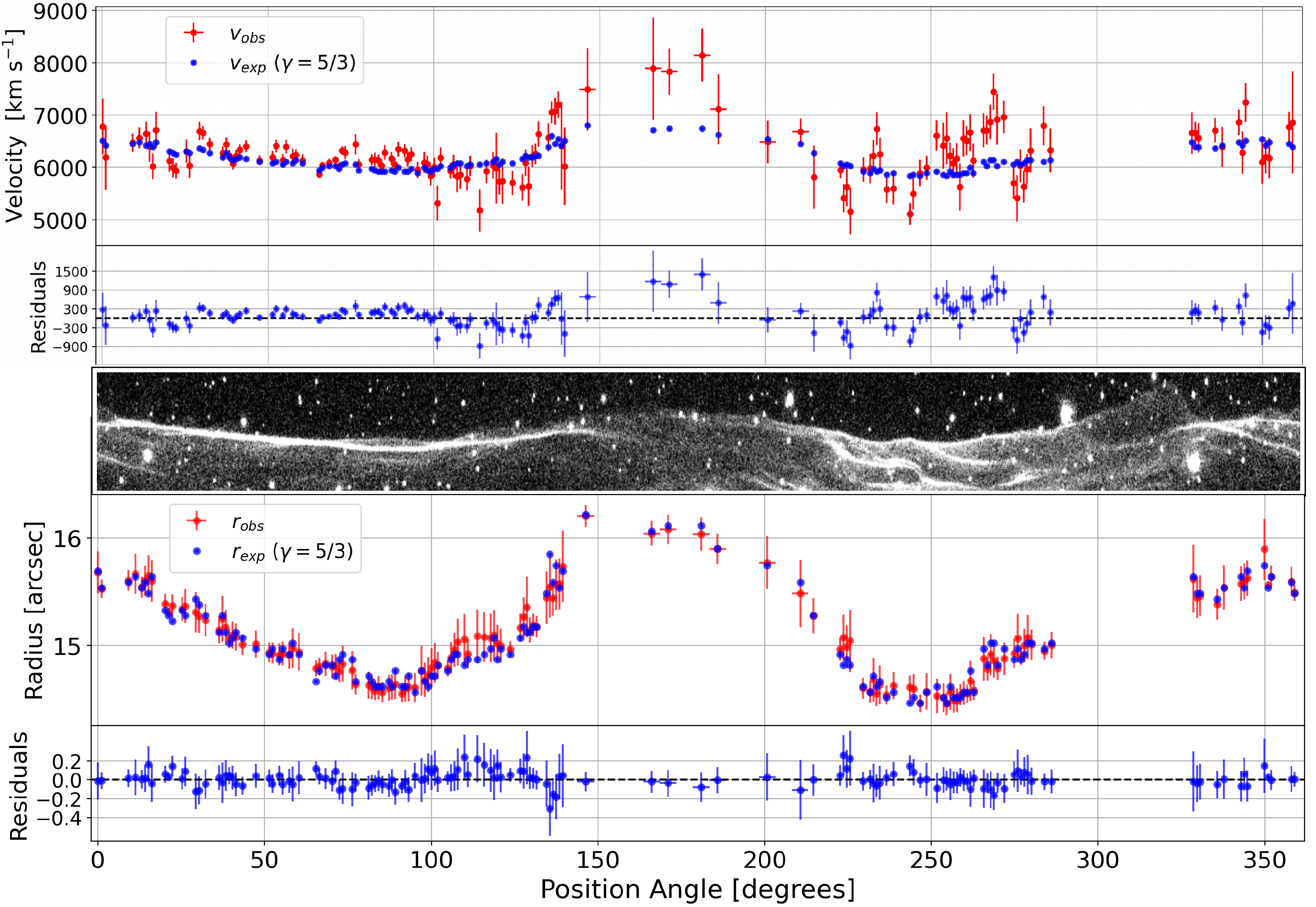}
        \caption{The measured values (red points) and best-fit results (blue points) from our MCMC run with a fixed explosion center for the FS velocity (top two panels) and radius (bottom two panels) along with residual plots.}
        \label{fig:residues}
    \end{figure*}

In figure~\ref{fig:residues} we compare the measured values of velocity (top panels) and radius (bottom panels) to the expected values based on the best-fit parameters obtained from the maximum likelihood MCMC model. Nearly everywhere, there is a close overlap of the expected model radii (shown in blue) with the observed values (red). The expected model velocities generally show a good match with the observed values for most of the regions, but are unable to follow the small-scale ($\sim$10$^\circ$) variations around the rim and fail to match the highest velocities seen in the faint southern rim.  The higher velocity residuals produce the significantly higher best-fit $\chi^2$ values for the velocity data alone compared to the radius data (see the last two columns of Table \ref{tab:MCMC_tab}). We discuss the radius/velocity comparison in more detail below (\S \ref{sec:rad_vel}). \par

Figure~\ref{fig:rho}a shows the best-fit ambient medium densities as a function of PA from the MCMC runs where age and distance are allowed to vary (blue points with uncertainties). As noted above there is a correlation between the distance and density in these fits and the plotted error bars include this correlated error. To better understand the intrinsic scatter in the densities, we re-ran the MCMC with $T$ and $D$ fixed at their best-fit values.  The density values and error bars from this run are over-plotted with orange points. The reduction in the error bars is quite significant, going from typical fractional errors of $12.9\%$ to $6.2\%$.  This demonstrates that the intrinsic random uncertainty in the density values is far smaller than the correlated uncertainty from age and distance. 

The best-fit age and distance derived using the different values of $\gamma_{\rm eff}$ are statistically the same. And, although the best-fit densities show the identical variation with PA, the ratio of best-fit densities is 0.63, with the $\gamma_{\rm eff} = 4/3$ fits yielding smaller ambient density values. The overall statistical quality of the fits is about the same. Since we do not have an independent measurement of the age or the ambient medium density, we recover an obvious but important conclusion from HHE15: with forward shock kinematic measurements alone it is not possible to discriminate between different shock compression factors arising from more or less efficient levels of DSA.

The mean ambient density surrounding \remnant for the $\gamma_{\rm eff} = 5/3$ case is  $6.35\times10^{-25}$ g cm$^{-3}$ and the maximum and minimum densities around the rim are factors of 1.3 and 0.59 times the mean value, respectively. The density variation around the rim is smooth and well-behaved.

For this set of MCMC runs we allow for the possibility that \remnosp's dynamical center (or equivalently its center of expansion) is not at the geometric center.  All of our measurements of shock position and velocity are local, that is, carried out using only the location and orientation (i.e., shock normal) of each specific azimuthal portion of the rim under consideration.  This allows us to calculate the shock radius on-the-fly within the MCMC calculation as the dynamical center changes; we assume that the shock velocities remain unchanged as the center moves around. The PA of the various apertures may change slightly, but these are purely labels and do not enter into the MCMC calculation.
This run yields optimized values for the center positions $\alpha_{\rm{dyn}}$ and $\delta_{\rm{dyn}}$ in the context of the hydrodynamical solution for the entire remnant.  Also, of course, the posterior provides confidence intervals on the optimized fitted $\alpha_{\rm{dyn}}$ and $\delta_{\rm{dyn}}$ values. 

As a specific validation of the MCMC code for fitting the dynamical center, we simulate a SNR evolving in an environment with a density asymmetry of a factor of 2 along the E-W axis. The simulated data are radius and velocity values for 90 azimuthal locations about an input dynamical center. The simulated uncertainties were drawn from the golden-sample error distributions for radius and velocity. The locus of the simulated radius values is close to circular at a center (geometric) some 17 pixels away from the dynamical center along the E-W axis. Applying the MCMC analysis to the simulated data, we recover the input dynamical center to $\sim 1$ pixel ($0.05^{\prime\prime}$) accuracy, and the input factor of 2 density gradient. This is a strong confirmation of the ability of the MCMC to recover the dynamical center in remnants evolving in environments with strong density gradients.

  Equipped with this validation, we run the MCMC for the 144-region golden sample. For the two additional parameters describing the explosion center ($\alpha_{\rm{dyn}}$ and $\delta_{\rm{dyn}}$), our MCMC recovers strongly peaked, approximately Gaussian posterior distributions for both with remarkably small uncertainties (roughly 1 {\em HST} pixel). The best-fit estimate of this new center, hereafter the {\em dynamical center}, is $\alpha_{\rm{dyn}}= $ $5^{\rm h}09^{\rm m}31^{\rm s}.16$ and $\delta_{\rm{dyn}}= $ $-67^{\circ}31^{\prime}17.1^{\prime\prime}$.  This is nearly identical to the geometric center; the locations are separated by only $0.03^{\prime\prime}$. Table \ref{tab:MCMC_tab} (bottom panel) shows the best-fit MCMC estimates for all the global parameters of this run along with the difference between the dynamical (fitted) and geometric centers. Figure~\ref{fig:mcmc_corner_fs}(b) shows the posterior distributions; we see the same covariances between $T, D$ and $\rho_i$ as observed in figure~\ref{fig:mcmc_corner_fs}(a). However, $\alpha_{\rm{dyn}}$ and $\delta_{\rm{dyn}}$ are uncorrelated with any of the other parameters, establishing that the dynamical center is independent of any of the global or local parameters. Table \ref{tab:MCMC_tab} also shows the $\chi^2$ contributions of the FS radius and velocity at the best-fit values. Note that when compared with the previous MCMC analysis with a fixed center, the reduction in total $\chi^2$ value with the addition of two parameters is not statistically significant. In other words, the dynamical center is statistically in agreement with the geometric center. As a verification test, we initiated the MCMC with a trial center several arcseconds from the geometric center, and, although the initial fit was quite poor, after convergence, the MCMC still yielded the same best-fit values for $\alpha_{\rm{dyn}} $ and $\delta_{\rm{dyn}}$. We discuss the astrophysical implications of the agreement in the dynamical and geometric centers of \remnant  in \S \ref{sec:kin_center}.   \par

Our derived ages and densities depend on the values of $M_{\rm{ej}}$ and $E_{51}$ assumed, although the effect is modest. Considering values of $E_{51}$ and $M_{\rm{ej}}$ between 1.0 and 1.4 yield ranges of  303--330 yr for age and 
 $3.8\times 10^{-25}$--$7.2\times 10^{-25}$ g cm$^{-3}$  for density. Younger ages go with higher densities and correspond to the lower mass-higher energy combination and vice versa. These follow from the characteristic scaling values (see equation~\ref{eqn:ch}).

\subsection{MCMC including the reverse-shock position}
\label{sec:energy_results}

For the previous MCMC runs, we fixed the explosion energy since we are unable to constrain this quantity (or the ejected mass)  with just information on the radius and velocity of the FS. However, this degeneracy is lifted when the RS position is included as an additional data point. To estimate the RS position we use the average radius of the [Fe XIV] $\lambda$5303  coronal line emission reported by \citet{seitenzahl+19}. These authors approximated the emission as a projected spherical shell with a radius of $2.86\pm 0.10$ pc. However, the [Fe XIV] $\lambda$5303 coronal line emission is just a proxy for the reverse shock position; it needs to be corrected to a smaller value using models for the post-shock increase in ionization and temperature.  \citet{seitenzahl+19} do this calculation assuming a core-envelope ejecta density profile (uniform density core surrounded by a $n=7$ power-law density envelope) in the context of the remnant's hydrodynamical evolution.  Their estimated location for the reverse shock is 2.74 pc, about 2.5\% to 3.9\% less than the radius of the coronal emission (given the range of results in their model, see Table 1 in \citealt{seitenzahl+19}). We run our MCMC analysis here adopting a RS radius of $11.30^{\prime\prime}$ (converted using an LMC distance of 50 kpc as these authors do) with an uncertainty of $0.41^{\prime\prime}$ which is a slight overestimate to account for the uncertainty in the modeling of the RS position. \par

\subsection{MCMC with a fitted dynamical center}
\label{sec:fitted_center}

    \begin{figure*}[t!]
        \centering
        \includegraphics[scale=0.4]{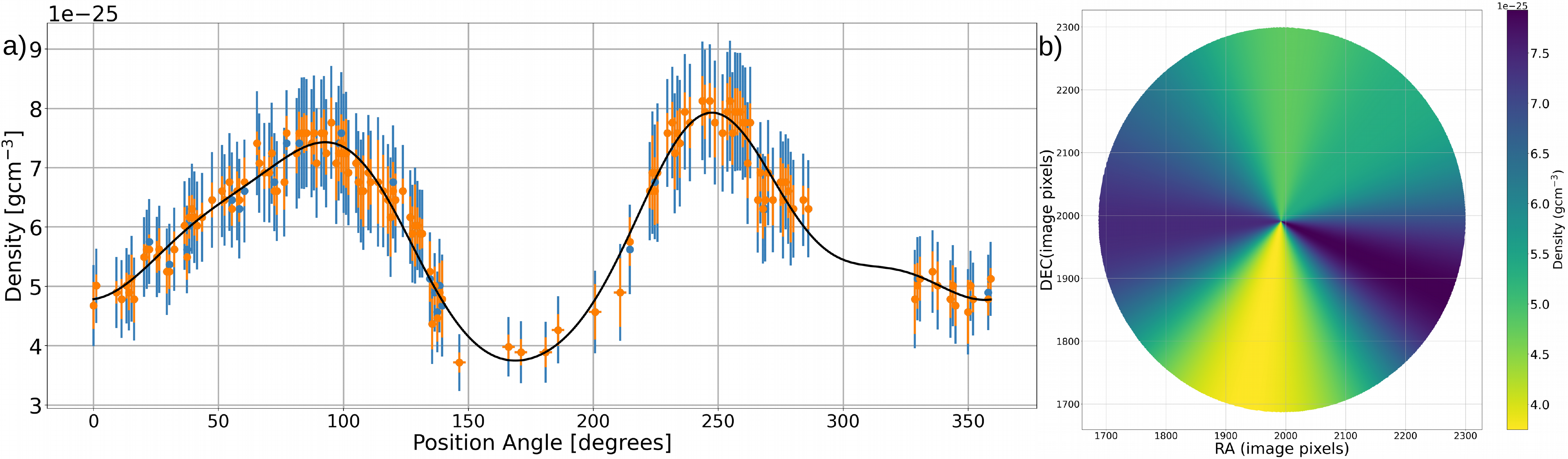}
        \caption{ a) MCMC estimates of the ISM density surrounding \remnant (for $\gamma_{\rm eff} = 5/3$) as a function of position angle. The blue points are the density values obtained when Age (T) and Distance (D) and density ($\rho_i$) are all allowed to vary, while the orange points are when T and D are fixed at their estimated values. The black curve shows the harmonic fit to the blue data points (see \S\ref{sec:densities}). b) A schematic visualization of the density map of \remnant obtained from the harmonic fit shown in panel a (see \S\ref{sec:densities}).}
        \label{fig:rho}
    \end{figure*}

Since we have only a single average data point for the RS radius, we correspondingly use one average data point each for the FS radius and velocity in our MCMC.  For this we compute the mean and  standard error on the mean for the FS radius and PM measurements of the golden sample (144 azimuthal positions); the computed values are $R_{\rm FS} = 15.0542 \pm 0.0341$ arcsec (which assumes the geometric center) and $S_{\rm b} = 5.2982 \pm 0.0352$ pixels (equivalent to a velocity of $6270 \pm 40$ km s$^{-1}$). The 3 available data points allow us to constrain 3 parameters: age, explosion energy (or mass), and a single global average density. The remnant's explosion center is fixed at the geometric value ($\alpha_{\rm{geo}}$ and $\delta_{\rm{geo}}$) and the LMC distance at   49.27 kpc.  When $E_{51}$ is being fitted, we set $M_{\rm{ej}}=1.4 M_{\odot}$; conversely, when $M_{\rm{ej}}$ is being fitted, we set $E_{51}=1.4$.  We also run two cases with different values of the effective adiabatic index: $\gamma_{\rm eff}=5/3$ and $4/3$.  As we will see, this physical parameter has a strong effect on the fitted explosion energy or ejecta mass.

\begin{figure*}[t!]
        \centering
        \includegraphics[scale=0.45]{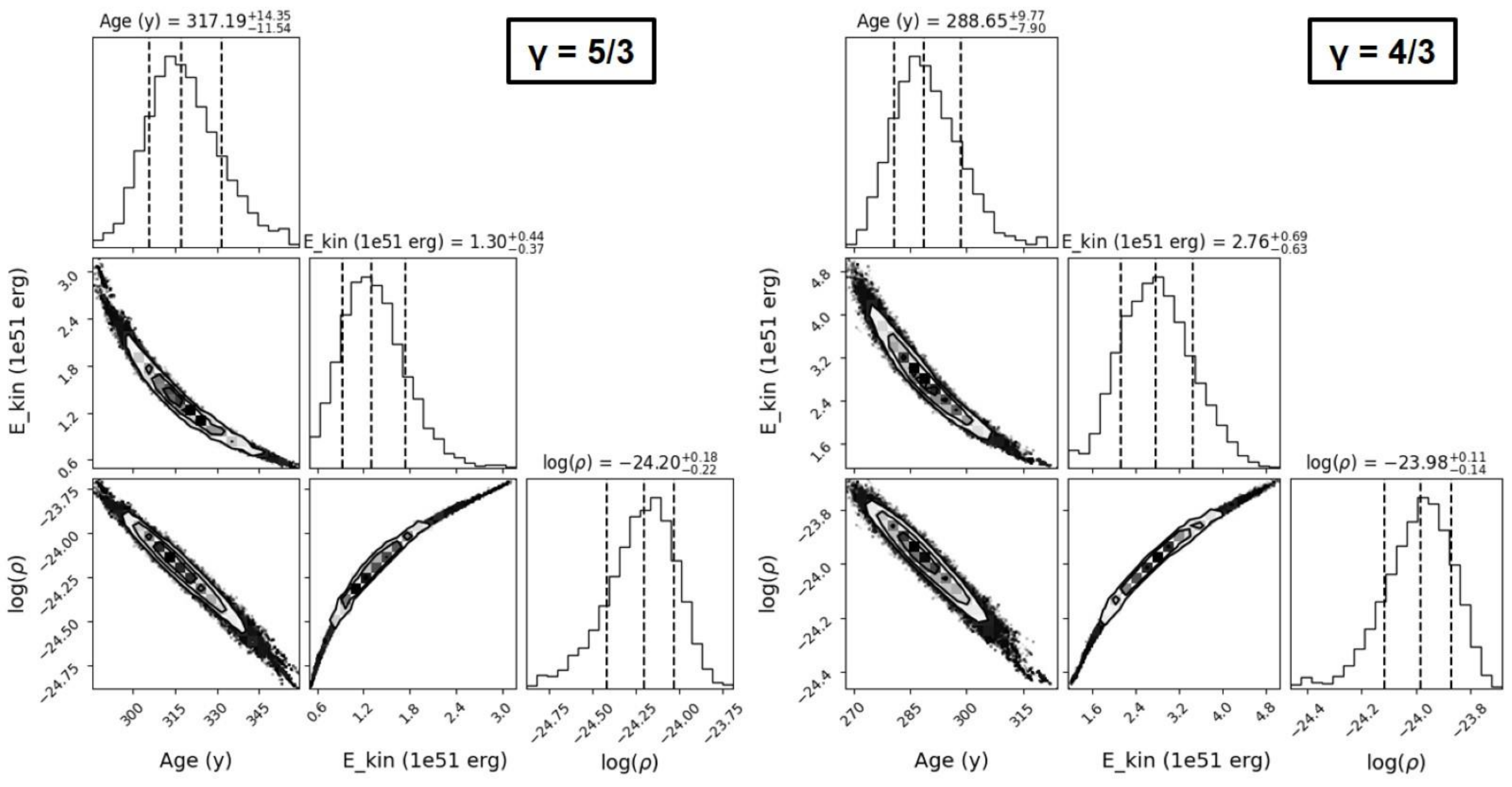}
        \caption{Corner plot showing the posterior distributions of the MCMC runs that fit for the explosion energy ($E_{51}$), for the two cases of $\gamma_{\rm eff} = 5/3, 4/3$ ((a) and (b) respectively).  The parameters are highly sensitive to $\gamma_{\rm eff}$, as reflected in their estimated values.}  
        \label{fig:mcmc_corner_rs}
    \end{figure*}

Figure~\ref{fig:mcmc_corner_rs} shows the corner plots for the two different $\gamma_{\rm eff}$ runs where the explosion energy is let free. The 1D posteriors for age and density are up to an order of magnitude wider than for the full FS MCMC runs alone (figure~\ref{fig:mcmc_corner_fs}) and are also somewhat asymmetric.  The 2D posteriors show significant correlations among all three fitted quantities with narrow, but very long, ``bananas'' reflecting the tight measurement uncertainties on FS radius and velocity and the much less constraining uncertainties on the RS radius. Still, the typical age constraints ($\sim \pm 10$ yr) are considerably better than our previous constraint from HHE15 ($\pm 40$ yr). 

\begin{table}[ht]
\centering
\caption{Estimates of Age, Ejecta Kinetic energy/mass and Density of SNR \remnant}
\setlength{\extrarowheight}{6pt} 
\begin{threeparttable}
% \color{red}\begin{tabular}{c c c c c}
\begin{tabular}{c c c c c}
\hline
\hline
$\gamma_{\rm eff}$ & $T$ (yr) & $E$ ($10^{51}$ erg) & $M_{\rm ej}$ ($M_{\odot}$)  & log($\rho$) \\ 
\hline
\multirow{2}{*}{5/3}  & $317.19\pm 12.95$ & $1.30\pm 0.41$ & $ 1.4^* $ & $-24.20\pm 0.20$ \ \ \\
& $329.7\pm 17.3$ & $1.4^*$ & $2.02\pm 0.85$ & $-24.23\pm 0.09$ \ \\
\multirow{2}{*}{4/3} & $288.65\pm \phantom{0}8.83$ & $2.76\pm 0.66$ & $ 1.4^* $ & $-23.98\pm 0.12$ \ \\
& $293.8\pm 10.6$ & $1.4^*$ & $0.82\pm 0.24$ & $-24.29\pm 0.03$ \  \\
\hline
\end{tabular}
\label{tab:mcmc_energy_results}
\begin{tablenotes}\footnotesize
\item {Note$-$} MCMC posteriors are obtained using single average values of FS radius and velocity and RS radius. Distance fixed at $  49.27$ kpc, center fixed at the geometric center. The other values kept fixed in the individual MCMC runs are indicated with $^*$. 
\end{tablenotes}
\end{threeparttable}
\end{table}

% \begin{table}[ht]
% \centering
% \caption{Estimates of Age, Ejecta Kinetic energy/Mass and Density of SNR \remnant}
% \setlength{\extrarowheight}{6pt} 
% \begin{threeparttable}
% \begin{tabular}{c c c c c}
% \hline
% \hline
% $\gamma_{\rm eff}$ & $T$ (yr) & $E$ ($10^{51}$ erg) & $M_{\rm ej}$ ($M_{\odot}$)  & log($\rho$) \\ 
% \hline
% \multirow{2}{*}{5/3} & $318.2\pm 12.0$ & $1.32\pm 0.38$ & $1.4^*$ & $-24.24\pm 0.18$ \ \\
% & $329.2\pm 17.1$ & $1.4^*$ & $1.92\pm 0.80$ & $-24.27\pm 0.09$ \ \\
% \multirow{2}{*}{4/3} & $288.9\pm \phantom{0}9.0$ & $2.83\pm 0.68$ & $1.4^*$ & $-24.01\pm 0.13$ \ \\
% & $293.7\pm 10.3$ & $1.4^*$ & $0.79\pm 0.22$ & $-24.33\pm 0.03$ \ \\
% \hline
% \end{tabular}
% \label{tab:mcmc_energy_results}
% \begin{tablenotes}\footnotesize
% \item {Note$-$} MCMC posteriors obtained using single average values of FS radius and velocity and RS radius. Distance fixed at $  49.27}$ kpc, center fixed at the geometric center. The other values kept fixed in the individual MCMC runs are indicated with $^*$. 
% \end{tablenotes}
% \end{threeparttable}
% \end{table}

All of the numerical results of these MCMC runs are shown in Table \ref{tab:mcmc_energy_results} for the two values of $\gamma_{\rm eff}$ used and two cases each where either $E_{51}$ or $M_{\rm{ej}}$ is allowed to be free. The uncertainties in the table have been symmetrized.

First, we note that the values of the global parameters are highly sensitive to the value of $\gamma_{\rm eff}$, as evident from the $\sim$30 yr difference in ages and factor of $\sim$2 differences in explosion energy or mass. Furthermore, the MCMC runs return broadly consistent values for the ratio of $E_{51}/M_{\rm{ej}}$ for each value of $\gamma_{\rm eff}$ given the uncertainties:  $\sim$0.8 for $\gamma_{\rm eff}=5/3 $ and  $\sim$1.8 for $\gamma_{\rm eff}=4/3$. However the effect on the best-fit values of allowing  $E_{51}$ or $M_{\rm{ej}}$ to vary is not entirely symmetric. In the case of varying $E_{51}$, the ambient density value is a factor of 1.7 times higher for the lower value of $\gamma_{\rm eff}$; when $M_{\rm{ej}}$ is allowed to vary instead, the ambient density value is a factor of 0.87 times less for the lower $\gamma_{\rm eff}$ value.  The dimensional scaling laws for radius, velocity, and age (equation~\ref{eqn:ch}) are not symmetric in $E_{51}$ and $M_{\rm{ej}}$, so differences in the detailed results from the MCMC runs are expected. 

The results for the two values of $\gamma_{\rm eff}$ reveal a very important difference.  The $\gamma_{\rm eff}=4/3$ case yields estimates of $E_{51}$ and $M_{\rm{ej}}$ that are much larger, or much smaller, respectively, than the typical expected values for SN Ia explosions, even considering the large uncertainties. On the other hand, for the $5/3$ case, the best-fit values for ejected mass and energy show excellent consistency with the highly-energetic, SN 1991T analog for the explosion of \remnant \citep{rest2008}. \par

\section{Discussion}
\label{sec:discussion}
In this section, we first compare and discuss the measured and fitted values of the basic kinematic quantities (radius and velocity) at the heart of our \remnant\ study. This leads to a deeper consideration of the underlying assumptions inherent in our hydro/MCMC formalism. Next, we compare our constraints on the dynamical quantities with existing estimates from the literature. The implications of our new dynamical center for the remnant are addressed briefly.  We discuss how our fits using the reverse shock radius, which allows constraints on the explosion energy or ejected mass, fit in the context of shock-modification due to diffusive shock acceleration.  We also present a graphical view of the variation of ambient medium  density in the vicinity of \remnosp.

\subsection{MCMC model fits and their validity}
\label{sec:rad_vel}

As noted above (\S~\ref{sec:fixed_center}) the FS radius values from our MCMC runs are in excellent agreement overall with the data (see figure~\ref{fig:residues} bottom panel). The FS velocity estimates on the other hand (top panel) show varying levels of agreement around the rim. 

To gain insight into the validity of our model approach, in the following, we examine specific sub-regions of the data, which stand out either due to the pattern of their residuals, their location around the rim, or due to other specific local details (in comparison with the general behavior). For these sub-regions, we rerun the MCMC using the fixed geometric center, $\gamma_{\rm eff} = 5/3$, and fixed values for $E_{51}$ and $M_{ej} (M_{\odot})$ of 1.4, i.e., similar to \S \ref{sec:fixed_center}. However, instead of fitting for the distance, we fix it at the MCMC best-fit value of $50.9$ kpc. \par

\textbf{Eastern rim:} In the eastern rim over a PA range of $28^{\circ}$ to $100^{\circ}$, our dynamical model agrees very well with both the radius and velocity data sets. The main shock of this arc, approximately $5$ pc long, is smooth (largely without dimples), sharp, and easily identified. Over the extent of the arc from north to south, the radius values fall from a high of 15.3$^{\prime\prime}$ to a low of 14.6$^{\prime\prime}$ while the velocities also fall from 6690 km s$^{-1}$ to 5840 km s$^{-1}$. The best-fit density values also show a variation, in this case increasing from a low value of  $5.2 \times 10^{-25} $ g cm$^{-3}$ at the northern end of the arc to  $7.8\times 10^{-25} $ g cm$^{-3}$ at the southern end. The variation of these three quantities is qualitatively consistent with expectations based on the observed ellipticity of  \remnosp, where the long axis of the rim is oriented close to the north-south direction. The modest level of model-data residuals for both the arc's radii and velocities show quantitatively that our dynamical model, including the assumption of constant density throughout the evolution, is a satisfactory approximation to the actual physical situation along the northeast rim.

\textbf{Eastern vs.\ western bright rim comparison:} Another way to check the validity of the dynamical model is to fit subsets of the data independently and compare the derived parameter values to those from the entire rim.  For just such a test we separately fitted the radii and velocities from a subset of apertures along the eastern bright rim (PA $\sim 80^{\circ}$) and a different subset from the western side (PA $\sim 250^{\circ}$), assuming the fixed geometric center. The fitted results are very close to those obtained using the full golden sample. The age values from the separate fits ( 311 and  325 yr) closely bracket the full sample value ( 316 yr), while the ambient density values were only slightly reduced ($-4$\% eastern rim) or increased ($+7$\% western rim). This excellent numerical agreement provides additional evidence supporting the validity of the dynamical modeling of the bright rims.  It suggests further that the underlying density distribution is fairly uniform across the middle of \remnant from one side to the other, a distance of $\sim$6.8 pc.

\textbf{Southern rim:} Now we turn to the velocity estimates from the MCMC along the southern rim (PA $146^{\circ}-186^{\circ}$) that show large differences compared to the data, with model expectations on average 800 km s$^{-1}$ too slow.  The radii fits in the southern rim, however, are consistent with the observations. This discrepancy suggests that the dynamical model is not accurately describing the observations in this part of the rim for a global best-fit age of 316 yr. Indeed we can obtain a good fit to both the velocity and radius data here if we just fit these five points separately in the MCMC.  In this case, we obtain a significantly lower age of 285 yr and density values that are reduced to $66\%$ of the global fit values. From this, we suggest that, unlike the bright rims, the faint southern rim is not well described by evolution through a constant density medium. Given that the kinematics of the eastern and western rims are separately consistent with a relatively uniform density across the middle of the remnant, the discrepancies with the velocity fit in the southern rim point to evolution through a varying density medium that starts out high near the explosion center and drops away with distance.  We leave the numerical investigation of this issue to a future study.

\textbf{Small-scale velocity fluctuations:} A final point to consider is the significant velocity structure on scales of $\sim$10$^\circ$ around the remnant's rim, which the MCMC analysis fails to reproduce. For example, there is a local peak in the FS velocity near PA of $\sim$136$^\circ$, 
a `W' like feature over a PA range of $215^{\circ}-260^{\circ}$, and another peak around PA of $\sim$270$^\circ$.  Neither the fitted velocities nor the ambient medium densities follow these relatively high spatial frequency variations.  The first local-peak feature lies in the southeastern  quadrant where the outermost rim (which is what we are tracking) transitions from a bright arc to a much fainter shock feature.
The other two features mentioned are in the western quadrant where there is a complex set of interior filaments and a number of dimples in the outer shock front.  We suggest these velocity fluctuations indicate the presence of modest density inhomogeneities in the environment, which have given rise to local variation in the shock velocities. When the shock encounters a local density variation there can be an appreciable and nearly immediate change in the shock velocity, but it takes some time for the change in shock speed to manifest as a significant change in radius. Thus, in a broad sense, the shock velocity is sensitive to the local ambient conditions at the forward shock, while the radius is an average, integrated measure of the ambient conditions through the course of the remnant's evolution.

\subsection{Comparison with results in the literature}
\label{sec:compare}

Here we compare our results to previously published work, focusing on a few specific studies that have given insights into the age \citep{rest2005} or the overall dynamical state of SNR \remnosp: \citet{badenes2008}, \citet{kosenko2014}, HHE15, and  \citet{seitenzahl+19}.  Results on the relevant dynamical quantities are given in Table~\ref{tab:litcomp}. While we use our results of the MCMC runs which fit for the explosion energy (using a single average FS radius and velocity plus RS radius, see \S~\ref{sec:energy_results}) to compare to the literature, we also refer to the MCMC results for the fixed energy case wherever relevant.  Since we use a different distance (49.27 kpc) than the other cases here, we present rough LMC distance scalings. When the distance to 50 kpc , then the inferred values from our MCMC analysis change in this manner: age increases by $<<$1 $\%$, and density decreases by $\sim$8$\%$. These are modest differences, and do not significantly influence the following discussion. 

\textbf{\citet{rest2005}}: Our age measurements are consistent with \citet{rest2005}, who use light echoes of the SNe event to determine an age of $400 \pm 120$ yr. The uncertainty quoted here comes from the standard deviation of age estimates from six separate light-echos associated with \remnosp, while their value of $400$ yr is calculated assuming the dust sheets scattering each echo are perpendicular to the line of sight (inclination angle of $0^{\circ}$). The authors state that the systematic uncertainty in the dust sheet inclination introduces the largest uncertainty in their age estimate.  If the inclination angle is reduced to $60^{\circ}$, the calculated age falls to 250 yr. Considering both the random and systematic uncertainties, the light-echo age estimate is  clearly consistent with our dynamically-determined one. We note that the assumption of a perpendicular  dust sheet inclination also overestimates the light-echo age measurement of SN1987A \citep{rest2005}. \par

\textbf{\citet{badenes2008}}: These authors take a sequence of spherically-symmetric SN Ia delayed-detonation \citep[DDT,][]{Khokhlov1991} explosion models from \citet{bravo+1996} and, using a 1D hydro code, evolve them through their interaction with a uniform ISM to an age of a few hundred years. The explosion models have an approximately exponential density profile and a modeled chemical composition profile that varies radially from a predominantly iron-rich center, to a zone containing intermediate mass elements (e.g., Si, S, Ar, Ca), and then an outermost oxygen-rich shell. These authors model the electron temperature assuming non-equilibration with ions and the nonequilibrium ionization states of the reverse-shocked ejecta to simulate the emergent X-ray spectrum.  

For the kinematic constraints, they use values for the FS available at the time. Their radius is the \textit{Chandra}-measured value of $15.1''$ ($R_{\rm FS} = 3.66$ pc), which is close to the average radius we measure of $15.2''$ with {\it HST}.  Our measurements, however, vary around the rim ranging over values of $14.5''-16.5''$  due to the ellipticity of the remnant. \citet{badenes2008} use an average FS velocity of $5350\pm 1750$ km s$^{-1}$ which comes from modeling the integrated broad Ly$\beta$ line width from {\it FUSE} observations of \remnant  \citep{ghavamian2007}.  In comparison the average FS velocity from our \ha PMss is $6270\pm 40$ km s$^{-1}$, plus we measure its variation around the rim.  

Given the large uncertainty on the FS velocity in the earlier work, their kinematic constraints on the age and ambient medium density are quite broad: ages from 250 yr to 600 yr and densities from $7\times 10^{-26}$ g cm$^{-3}$ to $3\times 10^{-24}$ g cm$^{-3}$ are allowed. The principal constraining power in this work comes from the photon energy centroids and fluxes of strong line emission complexes of oxygen, silicon, and iron in the well-observed X-ray spectrum. Combining the broad kinematic constraints and X-ray spectral measurements yields an acceptable solution that has an age of 400 yr, an ambient medium  density of  $1 \times 10^{-24}$ g cm$^{-3}$, and a strong preference for the specific DDTa explosion model. This explosion model has a kinetic energy of $E_{51} = 1.4$ and $M_{\rm ej} = 1.37\, M_\odot$ which is in good agreement with our estimate based purely on three kinematic measurements (see last column of table~\ref{tab:litcomp}).  On the other hand, though, their preferred age and ambient medium density are in strong  disagreement with our values, which are the properties that we measure most precisely from just the FS kinematics. Now that the kinematics of \remnant are so well constrained it would be useful to revisit the X-ray analysis of \citet{badenes2008}. \par

\textbf{\citet{kosenko2014}}: These authors present a library of models for young SNRs derived from multidimensional hydrodynamical simulations that include the effects of nonlinear DSA. They model the ejecta density profile as a $n=7$ power law and  assume the ambient medium to be uniform. In one of the applications presented in this study, they apply their model to \remnant\ using the FS radius and velocity measurements from \citet{ghavamian2007a} as above.  From this they estimate an age of $360 \pm 50$ y and an ambient medium density of $n_0 = 0.1-0.3$ cm$^{-3}$ ($\rho = m_{\rm p} n_0 = [1.7 - 5.0] \times 10^{-25} $ g cm$^{-3}$), assuming $E_{51}=1.0$ and $M_{\rm ej}=1.4\ M_\odot$. These values are fully consistent with our MCMC results (see the last paragraph of \S\ref{sec:fitted_center}). Given the limited set of observational constraints, they do not obtain any results on the efficiency of cosmic ray acceleration in this remnant.

%  Their age is consistent with our MCMC results while their density values fall somewhat below our best estimate of the mean density. }

\textbf{HHE15}: Our ejecta model and general approach to measuring the PMss are equivalent to that of HHE15; the longer time baseline between the \ha images of \remnosp\ in the new work here has allowed us to significantly reduce the uncertainties on our results. This is best noted in the case of the fixed mass and explosion energy case (see \S\ref{sec:fixed_center} or \S\ref{sec:fitted_center}) with a best fit age of $ 315.2 \pm 1.8$ yr, compared to HHE15's value of $310 \pm 40$ yr. Also, our mean density value of  $6.35 \times 10^{-25}$ g cm$^{-3}$ broadly agrees with their value of $5.8 \times 10^{-25}$ g cm$^{-3}$, when the different assumed LMC distances are considered. Meanwhile, the uncertainties drop significantly to $\sim$1\% ($\sim$12\% per aperture) from $\sim$30\% (HHE15). Additionally, HHE15 estimate the remnant's age using FS radius and velocity values averaged over all PA. Using the hydro model and the  measured average radius and velocity, they map out the allowed parameter space for the age and ambient density using  $\chi^2$ as the figure-of-merit function. In our work, we are able to improve on this by using 144 radius and velocity measurements from apertures around the rim in an MCMC analysis, which naturally accounts for the variability in the FS radius and velocity due to the remnant's ellipticity, and provides a very precise global age for the remnant along with densities associated with each aperture. \par

\textbf{\citet{seitenzahl+19}}: These authors model the hydrodynamic evolution and ionization structure of an SN Ia remnant using a uniform core and $n=7$ envelope density profile for the ejecta expanding into a uniform density ISM. They assume an age of $310$ yr (from HHE15) and choose a uniform ambient density value of 0.4 amu cm$^{-3}$ ($6.6 \times 10^{-25}$ g cm$^{-3}$) to match the FS radius and velocity for their preferred values of $M_{\rm{ej}} = 1\, M_{\odot}$ and $E_{51} = 1.5$.  Corrected for the LMC distance, this is broadly consistent with our average density of  $7.2 \times 10^{-25}$ g cm$^{-3}$ determined for the same mass and a similar energy ($E_{51} = 1.4$) (see \S\ref{sec:fitted_center}). Although the exponential ejecta density profile we use is different from their core-envelope one, figure~10 in HHE15 shows that the age and ambient medium density derived from FS kinematics are similar for the exponential and $n = 7$ power-law profiles.

Our approach to constrain the mass and energy of the explosion is to use the kinematics of the FS and RS alone.  This leads to a discrepancy with \citet{seitenzahl+19} who are forced to a higher energy ($E_{51} = 1.5$) and lower mass ($M_{\rm{ej}} = 1\, M_{\odot}$) than our results in order to produce sufficient [Fe XIV] $\lambda$5303 emission from their model. To be more specific, we estimate $E_{51} =  0.91 \pm 0.30$ and $\rho = 0.24$ amu cm$^{-3}$ from our MCMC analysis and kinematic data when assuming their distance, mass value and RS location. The key issues we see that need to be investigated to resolve it include how collisionless heating of electrons at the reverse shock is modeled, assumptions about the profiles of density and elemental composition through the ejecta, and the role of variations in the ambient medium density (our analysis show a factor of two density variation around the rim).
\par

\begin{deluxetable*}{c c c c c c}[ht]
\tablecolumns{6}
\tablewidth{0pc} 
\tabletypesize{\scriptsize}
\tablecaption{Comparing estimates of dynamical quantities of SNR \remnant from literature}
 \tablehead{
 \colhead{} &
  \colhead{Badenes et al. (2008) } & \colhead{Kosenko et al. (2014)} & \colhead{HHE15}
    & \colhead{Seitenzahl et al. (2019) }
    & \colhead{This paper$^a$}} 
  \startdata
Ejecta profile & DDTa & power-law$^b$ & exponential & core-envelope$^c$ & exponential \\
Age (yr)  & 400 & 360 $\pm$ 50  & 310 $\pm$ 40 & 310  & 318 $\pm$ 13  \\
Distance (kpc) & 50 & 50 & 50 & 50 & 49.27 \\
Density (gcm$^{-3}$) &  $1\times 10^{-24}$   & $1.7-5.0 \times 10^{-25}$ & $5.8 \times 10^{-25}$ & $6.6 \times 10^{-25}$    & $ 6.24 \times 10^{-25}$ \\
E$_{51}$  & 1.4  & 1.0 & 1.4 & 1.5 & $ 1.30 \pm 0.41$        \\
$M_{\rm{ej}} (M_{\odot}$) & 1.37  & 1.4  & 1.4 & 1.0 & 1.4    \\
\vspace{-0.2cm}
 \enddata
 \tablecomments{$^a$Results shown for MCMC posteriors for the nominal case of $\gamma_{\rm eff} = 5/3$, where we fit for the explosion energy using a single average FS radius and velocity and RS radius. Distance is fixed. $^b$ with  $n=7$. $^c$ uniform density core with $n=7$ envelope.}
\label{tab:litcomp}
\end{deluxetable*}

\subsection{Ambient density spatial variation}
\label{sec:densities}
Our model determines the ambient densities around the rim, assuming that each region is evolving independently of the others, with a constant value throughout the course of the remnant's evolution.  Since we are viewing the FS generally edge-on, our radius and PM measurements are therefore being taken in the plane perpendicular to the line-of-sight (i.e., in the plane of the sky). Spectroscopy of Balmer-dominated shocks show that inclination angles are typically $>$80$^\circ$ with  brighter parts of the rim even  closer to 90$^\circ$ (e.g., \citealt{nikolic2013}).  In the following, it is important to recognize that the Balmer shocks we measure are sampling only a small portion of the ambient medium  surrounding \remnosp.

Our MCMC density estimates and their uncertainties indicate a relatively smooth, approximately sinusoidal pattern (see figure~\ref{fig:rho}) as a function of PA. The density gradient around the rim varies by a factor of 2.3 between the most and least dense regions, with a mean value of  $6.35 \times 10^{-25}$ g cm$^{-3}$ (0.38 amu cm$^{-3}$) for the typical case. \par 

In order to generate a visualization of the ambient medium density in the environment of \remnosp, we fit the densities, $\rho(\theta)$, shown in blue in figure~\ref{fig:rho}a, to a smooth (5th-order harmonic) function in PA, $\theta$, given by
$$
\rho(\theta) = \sum_{n=0}^5 a_n \sin(n(\theta-\phi_n)) \ .
$$
This function, as figure~\ref{fig:rho}a shows, produces good agreement for the locations and values of the two density minima and maxima as well as most of the large-scale asymmetries. The map of the ambient density in the plane of the sky (as shown in figure~\ref{fig:rho}b) is produced by plotting at each PA a single color from the center to the edge, in accordance with the constant density evolution assumed in our model. Near the center, the spatial variation in the density is unphysical, varying rapidly with PA and multivalued at the precise center.  This is a limitation of our simple 1D hydro model approach.

Now we attempt to synthesize the preceding discussion, as well as the comparison of fits from the eastern, western, and southern rims done in \S6.1, into a qualitative but physically-plausible picture of the ambient medium in the vicinity of \remnosp. One could convert the map in figure~\ref{fig:rho}b into a physical one by assigning a specific value (e.g., the average around the rim) to the density at the center.  In this case, for nearly all the radial spokes, the density would have to vary with radius by increasing (decreasing) outward in the E-W (N-S) direction to match the observed variation of radius and velocity.  Although physically realistic, this density map would not be very plausible, since it introduces a point-of-symmetry in the ambient medium centered on the remnant.  A more refined approach would take into account the symmetries apparent in the density field map, such as the east-west (100$^\circ$--250$^\circ$) alignment of the density maxima and north-south (165$^\circ$--320$^\circ$) 
alignment of the density minima. Utilizing such density maps would require multi-dimensional hydro modeling, significantly increasing the computational difficulty of a comparison to the radius and velocity data. We note, however, that our density estimates are the first step towards obtaining a more `data-informed' choice for the density map. \par 

 The preceding discussion was based on our baseline model which assumes that the remnant's ellipticity comes from a varying ambient medium density. 
% Our baseline model produces a non-spherical remnant taking the ambient medium density as the local variable.
However, a non-spherical SNR can also be achieved by modeling an asymmetric explosion energy in a uniform ambient medium. We consider this by running a MCMC for a uniform ambient density of $6.35 \times 10^{-25}$ g cm$^{-3}$ (the average of the golden sample densities for the baseline case), allowing the explosion energy to vary around the rim. The fit is as good as the baseline model and the posterior distributions for the age are equivalent. The explosion energy values range from $E_{\rm 51} = 1.20 - 1.79$ with a mean and RMS of $1.42$ and $0.16$ respectively. The higher energy values correspond to larger radii and vice versa. This analysis reveals that allowing the density or energy to vary around the rim can equally well describe the variation in the FS radius and velocity. Notably, for the energy-varying case, we find that the fitted center is shifted by $0.27^{\prime\prime}$ south-east of the dynamical center from the baseline model. This value is included as a systematic uncertainty in the position of the dynamical center, which we turn to next.  

\subsection{Dynamical center}
\label{sec:kin_center}
The remnant's geometric center is obtained by fitting an ellipse to the FS radii.  The remnant's dynamical center (explosion site) is constrained by our 1D hydro MCMC runs  (for the baseline model,) fitting the FS positions and PMss from all around the rim while the center position is allowed to be free. The fits (\S~\ref{sec:fitted_center}) find the remnant's dynamical center to be at $5^{\rm h}09^{\rm m}31^{\rm s}.16$, $-67^{\circ}31^{\prime}17.1^{\prime\prime}$, which is very close to the remnant's geometric center (within 0.03$^{\prime\prime}$). The 3$\sigma$ uncertainty in the position of our dynamical center $0.15''$ in right ascension and $0.19''$ in declination. \citet{schaefer2012}'s center differs from ours: it is some  $1.19^{\prime\prime}$ ($\sim 5\sigma$) away\footnote{ Our WCS solution differs from that of \citet{schaefer2012}. The value of $1.19^{\prime\prime}$ is after accounting for this correction.} toward the southwest.
These authors had applied a correction to their geometric center, shifting it 1.39$^{\prime\prime}$ to the southwest, writing that ``the expansion has recently slowed down'' in this quadrant.  Although there is evidence in this quadrant for an excess of X-ray \citep{warren2004}, FUV \citep{ghavamian2007a} and  24-$\mu$m IR emission \citep{Borkowski2006}, as well as a set of interior \ha filaments (figure~\ref{fig:0509}), our work shows that the dynamical center is not significantly offset from the geometric center.  This is a consequence of the symmetry of the ambient density (see figure \ref{fig:rho}), and the lack of a strong density gradient across the diameter (which can produce a significant shift in the dynamical center (see \S \ref{sec:fitted_center}). We conclude that the observed asymmetries in the remnant's emission toward the southwest are due to density enhancements out of the plane of the sky that the forward shock has encountered during later phases of its evolution. Thus an offset of the explosion center in the plane-of-the-sky frame is not necessary to explain the observations. \par 

   \begin{figure*}[t!]
        \centering
        \includegraphics[scale=0.64]{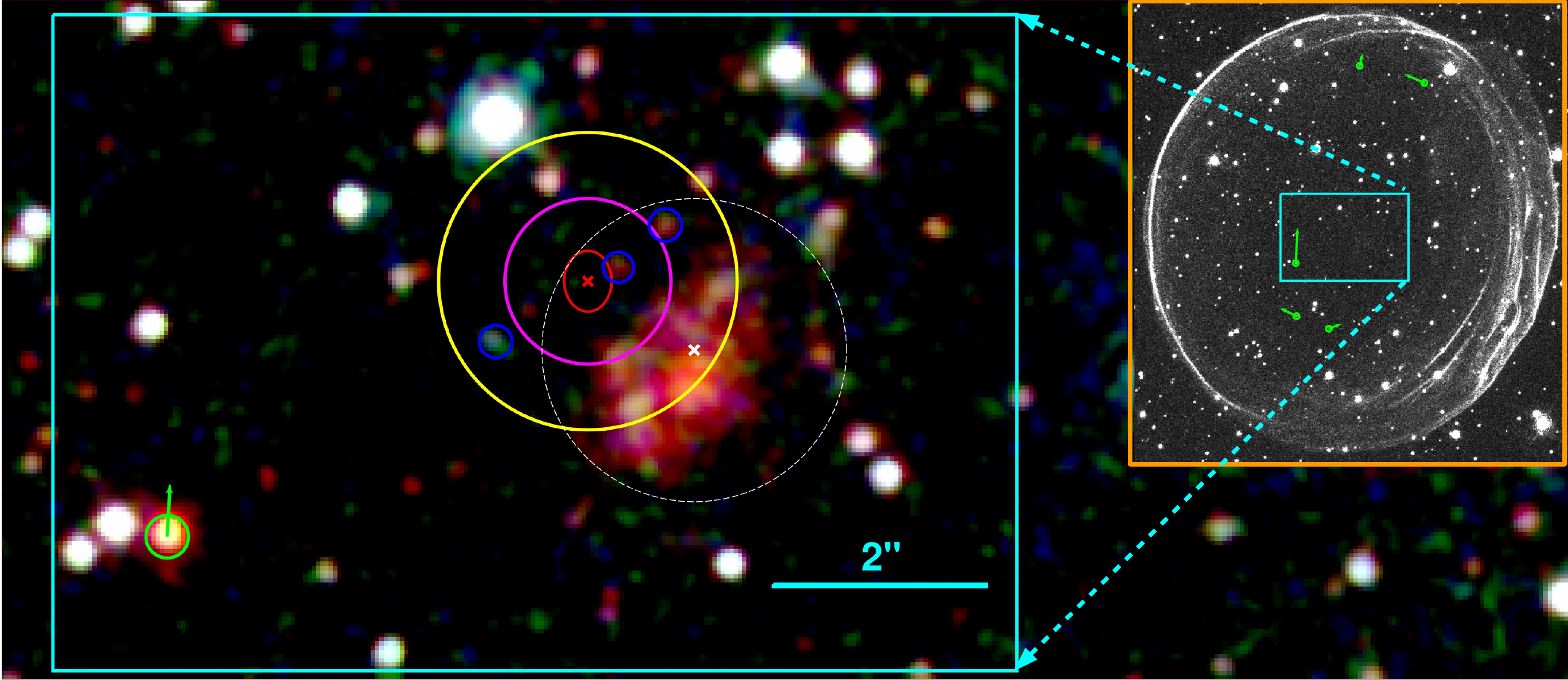}
        \caption{Three-color composite image of the central regions of SNR \remnant. The red cross and ellipse mark the dynamical center and its total measurement uncertainty. The magenta and yellow circles respectively show the search radius for main-sequence and white dwarf companion candidates. The white cross and the dotted circle mark the explosion center and error circle of \citet{schaefer2012}. The dark blue circles show 3 previously unidentified companion candidates that lie within the yellow circle. On the right, within the orange box, we show the epoch-1 H$\alpha$ image from 2006. In this image, we mark the 5 stars in green circles that have significant proper motion in the 10 year time frame between the epoch-1 and epoch-2 H$\alpha$ images. The magnitude (exaggerated for clarity, not to scale) and direction of proper motion are marked with green vectors. The star with the largest proper motion is the red star closest to the remnant's center moving at $0.01^{\prime\prime}$ yr$^{-1}$. The keen reader will note that the location of the red star in the composite image (from 2010) is slightly offset from the location of the green circle (which marks the star's location in 2006). This is consistent with the magnitude and direction of the red star's proper motion.   }  
        \label{fig:exp_site}
    \end{figure*}

Thanks to failed searches for left-over progenitor companions, the evidence against the single-degenerate (SD) scenario has been steadily building over the past few years. \citet{schaefer2012} rule out the SD channel basing their arguments on a lack of evidence for viable candidates near the vicinity of their center up to a V-band magnitude of $26.9$. In a separate analysis, \citet{Litke2017} also concur with this picture, using the color-magnitude diagrams (CMDs) of all stars within $3''$ of their geometric center (which is $0.08''$ away from our kinematic center). The authors find that none of their companion candidates show any overlap with the expected CMDs of post-impact evolutionary tracks of He and main-sequence (MS) companion stars \citep{Pan2014}. These authors, however, limit their search to a narrow range of stellar types (MS or He star) for the progenitor companion candidates, whereas the SD channel accommodates a broader range, with additional possibilities coming from the mode and rate of mass transfer between the progenitor and its companion. In this context, \citet{DiStefano2011} introduce the spin-up/spin-down scenario, where one can expect a significant delay ($10^5-10^9$ y) between the time the white dwarf (WD) spins up with sufficient accreted mass to achieve thermonuclear ignition, and the time it spins down to actually explode. This gives the companion star enough time to evolve significantly and thereby become fainter. Another alternative is the possibility of M-dwarf companion candidates which achieve mass transfer via magnetically confined accretion \citep{Wheeler2012}. Both these scenarios predict companion candidates that are much fainter than the detection threshold of existing data. \par

In recent years, a promising model for Ia explosions is the `helium-ignited violent merger' scenario \citep{Guillochon2010,Raskin2012,Pakmor2013}. This model, also termed the ``dynamically-driven double-degenerate double-detonation'' or D$^6$ model, posits that a Ia explosion is triggered in a double-degenerate (2 WDs) system coalescing during an unstable mass transfer phase. The donor WD undergoes a He-detonation on its surface during this time, which triggers a runaway thermonuclear explosion in the accreting C/O WD via the double-detonation channel. An exciting consequence of the D$^6$ model is the possibility of a surviving, runaway companion with a kick velocity of $>1000$ km s$^{-1}$. Support for this picture comes from the discovery of runaway WD candidates in the Milky Way by \citet{Shen2018b}, one of which can be traced back to a supernova remnant. This model provides another avenue for a surviving companion star and further reopening the search for a WD progenitor companion for \remnosp. \par 

 Figure \ref{fig:exp_site} shows the three-color HST composite image of \remnant with our constraints on the explosion center. The composite image is produced by combining the existing observations of \remnant from HST programs 12326 (PI: Knoll) and 13282 (PI: Chu) in the F475W (seen in blue), F555W (green), and F814W (red) filters. The red cross and the red ellipse ($0.22^{\prime\prime} \times 0.29^{\prime\prime}$) surrounding it mark our dynamical center and its measurement uncertainty combining the density-varying and the energy-varying results. The magenta error circle with radius $0.78^{\prime\prime}$ shows the search radius for a MS companion candidate. This value is obtained from the root-sum-square combination of the $3 \sigma$ dynamical center positional uncertainty and the maximum kick velocity of a MS companion star of $510$ km s$^{-1}$ \citep{canal2001} which translates to $0.69^{\prime\prime}$ at a distance of $49.27$ kpc in 316 yrs. The concentric yellow error circle of radius $1.40^{\prime\prime}$, obtained from a similar calculation, shows the search radius for WD companion candidates for a maximum kick velocity of $1000$ km s$^{-1}$ ($1.35^{\prime\prime}$). The white cross and the dotted circle show the explosion center and error circle from \citet{schaefer2012}. We find 3 previously unidentified candidates in dark blue circles, that lie within the yellow error circle. Although 2 of these candidates also fall within the error circle of \citet{schaefer2012}, these stars were only identified because of the deeper F814W observations, which were unavailable during their study.

 Finally, using the epoch-1 and epoch-2 H$\alpha$ images, we identify stellar candidates that show significant PM in the time frame between the two epochs. To do this, we use the IRAF \texttt{daofind} algorithm to map the locations of all point sources within a $11^{\prime\prime} \times 12.5 ^{\prime\prime}$ elliptical region within the remnant's interior in both the H$\alpha$ frames. Among these, we pick out 173 candidates whose peak brightness is $\ge 4.5\sigma$ above the background, and fit the root-sum-square distribution of the radial offsets between the two epochs of these candidates with a Rice distribution. Of these stars, we find that there are 5 candidates that have a $\ge3\sigma$ shift in their positions. A $3\sigma$ shift corresponds to $0.66$ HST pixels or $0.0033^{\prime\prime}$ yr$^{-1}$, which is $770$ km s$^{-1}$ at the LMC, a value that is in the range of potential kick velocities.  In Figure \ref{fig:exp_site} (orange box on the right), we show the epoch-1 H$\alpha$ image with the 5 candidates, along with arrows that show the direction of their PM. None of these stars appear to be moving radially away from the dynamical center, although it is interesting to note that the star with the maximum PM is the red star, identified as star M by \citet{schaefer2012}, closest to the dynamical center at a distance of $4.6^{\prime\prime}$. This star has shifted by 2 HST pixels in the time frame between the two epochs, which translates to a PM of $0.01^{\prime\prime}$ yr$^{-1}$, or $\sim 2300$ km s$^{-1}$ if in the LMC. 

 In this paper, our main focus is to determine an accurate and unbiased explosion center and its sources of uncertainty. With this first robust measurement of the dynamical center, the location of any potential progenitor companion is now dominated by astrophysics, rather than a measurement error in the center. To ascertain the nature of the faint sources detected, deeper observations must be pursued.

\subsection{Explosion Energy}
\label{sec:exp_en_discussion}

We have discussed above our method for constraining the explosion energy (\S~\ref{sec:energy_results}) and compared our results to previously published works (see \S~\ref{sec:compare}). In the present work, the addition of the RS location as a measured quantity has allowed us to constrain the explosion energy (or the ejecta mass) for the very first time from the kinematics of the FS and RS alone. Even considering the large uncertainties, our results are significant, as they provide an important step forward as the field moves from a model-driven to a more data-driven approach to understanding the physics of Ia explosions. Most of the previously published works, in contrast, make use of the complicated emission properties of the ejecta gas (as opposed to our use of the cleaner kinematic measurements) to infer the explosion energy and ejected mass. For the nominal $\gamma_{\rm eff} = 5/3$ case, we find a large inferred explosion energy for an assumed large ejected mass ($1.4\ M_\odot$). This is in accord with the spectroscopic sub-typing of \remnant as a luminous energetic SN1991T type SN from its light-echo spectrum \citep{rest2008}.  Our work provides an independent verification of this hypothesis through the remnant's kinematic properties.\par

Our estimates for the energy are also strongly influenced by the effective adiabatic index $\gamma_{\rm eff}$ at the FS, complicating the picture just described, but, on the other hand, giving us a window into the efficiency of CR acceleration in the Balmer shocks of \remnosp. Nonlinear diffusive shock acceleration introduces modifications to the hydrodynamical evolution of young SNRs \citep[e.g.,][]{decourchelle+00} producing an increased compression factor at the shock and a reduction of the spacing between the FS and RS.  These effects are clearly present in our models (see figure~\ref{fig:shock}). To match the observational constraints on the locations of both the forward and reverse shocks, the MCMC hydro model fits need to adjust an additional parameter beyond just the age and density, like the explosion energy or ejected mass. We choose to vary one of these quantities, while exploring a limited set of $\gamma_{\rm eff}$ values. At one limit ($\gamma_{\rm eff} = 5/3$, compression factor of 4) we obtain a minimum value for the explosion energy; as we reduce $\gamma_{\rm eff}$ the compression factor increases as does the inferred explosion energy.\par

Two previous studies by some of the current co-authors \citep[HHE15 and][]{hovey+18} have used \HST imaging observations to address the efficiency of CR acceleration. HHE15, specifically, rule out values of $\gamma_{\rm eff}< 1.2$ by coupling their dynamical results on the ambient medium density surrounding \remnant  with the inferred neutral hydrogen density from the H$\alpha$ surface brightness (see their figure 11). In \citet{hovey+18} shock velocities from \HST PM measurements of both \remnant and \remtwo\ are compared to H$\alpha$ broad-line widths.  Efficient shock acceleration of cosmic rays would result in less thermal heating of the ions producing lower broad line widths. This study found no evidence for lower line widths, setting an upper limit of 7\% for the CR acceleration efficiency at the FS of these two Balmer-dominated SNRs.\par 

In our study, for the $\gamma_{\rm eff} = 4/3$ case the MCMC analysis yields a best-fit explosion energy of  $E_{51} = 2.76\pm 0.66$ for fixed mass value of $M_{\rm ej} =1.4\ M_\odot$. This value is significantly higher than for the nominal $\gamma_{\rm eff}= 5/3$ case and is approximately 2$\sigma$ higher than the traditional ``high'' value of $E_{51}= 1.4$ for SN Ia explosions. Additionally, we obtain a best fit value of  $M_{\rm ej} = 0.82\pm 0.24 \ M_\odot$ when allowing the ejecta mass  to be fitted for a fixed energy value (1.4).  In either case, whether energy or mass is free, an unrealistic set of global values for the explosion properties results. \par 

 In a separate analysis, we compare our ambient density values with the average post-shock hydrogen number density from \citet{williams2011}. These authors fit their mid-IR spectra with dust-heating and sputtering models to derive a post-shock density of 0.59 cm$^{-3}$. Using the average compression factors for the $\gamma_{\rm eff} = 5/3$ case from Table~3 of HHE15, this translates to a pre-shock density of $4.77 \times 10^{-25}$ g cm$^{-3}$, a value that is only $1.3\sigma$ away from our value of $6.24 \times 10^{-25}$ g cm$^{-3}$. On the other hand  the $\gamma_{\rm eff} = 4/3$ case results in a pre-shock density value that is  $> 10\sigma$ away. Our dynamical model for the $\gamma_{\rm eff} = 5/3$ case is, hence, consistent with an independent density measurement from \textit{Spitzer} mid-IR observations.  \par

We conclude, from both the unrealistic set of global parameter values, and the large disparity with the density results from \citet{williams2011}, that the $\gamma_{\rm eff} = 4/3$ case is not a valid solution and can be ruled out, even though it describes the FS and RS kinematics in \remnosp . This provides additional evidence that cosmic ray acceleration is not highly efficient in this SNR. Further constraints would be possible with more precise measurements of the reverse shock. \par

\section{Conclusions}
\label{sec:conclusions}

In this work, we have focused exclusively on obtaining dynamical constraints on the explosion properties and environment of the young LMC SNR \remnant purely from kinematic measurements of the forward and reverse shocks.  The key measurements come from the {\it Hubble Space Telescope} whose state-of-the-art resolution allows for a detailed and comprehensive view of the forward shock of this remnant that lies some 50 kpc away. The SNR's forward shock emits only in the light of H$\alpha$, which we image using the ACS/WFC and the F658N filter.  The thickness of the shocked emission zone is unresolved even by \HST allowing highly precise measurements of the shock location and its PM at a large number of azimuthal positions (as many as 231, most $1^{\circ}$ wide) around nearly the entire elliptical rim. For the expansion, the time baseline of $\sim 10$ years between the two epochs of observation means that the rim has expanded radially by approximately 5 pixels (0.25$^{\prime\prime}$), resulting in a clear separation of the shocks in the two epochs. The mean 1-$\sigma$ statistical uncertainties on our FS measurements are 0.036 pc (1.0\%) and 310 km s$^{-1}$ (4.9\%) assuming the current best LMC distance estimate of   49.27 kpc.  The uncertainty on the distance, $\pm 1$ kpc (2\%), is of the same magnitude and is included as a Gaussian informed prior during our Bayesian analysis.

Another key feature of our work is the Markov chain Monte Carlo analysis we have implemented. At the core of the analysis is a 1D spherically symmetric hydro code that we use to produce similarity solutions for the evolution of shock radius and velocity as a function of time. This is done for two values of $\gamma_{\rm eff}$, $5/3$, and  $4/3$, to approximate the effects of diffusive-shock acceleration by modifying the compression factor at the FS.  We assume ejecta with an exponential density profile expanding into a uniform density ambient medium.  From the measured forward shock location and expansion speeds, the MCMC provides posterior distributions for the age and dynamical center of the remnant (global parameters), and the ambient medium density values for each of the 144 (``golden sample'') azimuthal apertures.

For an assumed ejecta mass of $1.4 M_{\odot}$ and energy of $1.4 \times 10^{51}$ erg, our best measurement of \remnosp's age is $317\pm 2$ yr. The density measurements vary around the rim in a remarkably orderly manner: the density peaks at the eastern and western rims (near PA $95^{\circ}$ and $245^{\circ}$) at roughly the same value of $7\times 10^{-25}$ g cm$^{-3}$ and reaches a minimum in the south (near $180^{\circ}$) at a density value 2.3 times less.

Because we determine the FS location and expansion speed based on local measurements at the shock front, we can fit for the remnant's dynamical center based on the kinematics of the FS around the rim. This is done by calculating the shock radius of the FS at each aperture on-the-fly during each step of the MCMC analysis. Our best-fit `dynamical' center is fully consistent with the geometric center, which is located at $5^{\rm h}09^{\rm m}31^{\rm s}.16$, $-67^{\circ}31^{\prime}17.1^{\prime\prime}$ with an uncertainty of $\lesssim 0.2''$ in each direction. This is the first robust measurement of the explosion center that takes advantage of all the available kinematic information on the FS, in the context of a dynamical solution.   Within a $1.4^{\prime\prime}$ search radius around the dynamical center, that accommodates for stars moving up to a kick velocity of $1000$ km s$^{-1}$, we find 3 previously unidentified faint sources. Assessing the nature of these sources requires deeper observations of the central regions of \remnant. Additionally, using the H$\alpha$ epoch-1 and epoch-2 images, we identify stars moving with a significant PM in the time-frame between the epochs. This analysis is sensitive to LMC stars moving up to at 770 km s$^{-1}$ or higher. The only star with a significant PM (as seen in H$\alpha$) in the vicinity of our center is the red star some $4.6^{\prime\prime}$ away, moving nearly upward (north) at a PM of $0.01^{\prime\prime}$ yr$^{-1}$. We find no evidence for any high-speed companion candidates that lie are moving radially from the dynamical center.  \par

The final significant result from our MCMC analysis is the measurement of the explosion energy, which we obtain by using the average values of the FS kinematic data and the RS location from \citet{seitenzahl+19}. Our best-fit explosion energy for a $1.4 M_{\odot}$ ejecta mass and the standard assumption of a $\gamma = 5/3$ gas is $E=(1.30\pm0.41)\times10^{51}\, \text{erg}$, supporting the spectroscopic sub-typing of SNR \remnant as a energetic 1991T-type explosion \citep{rest2008}. The cosmic-ray modified dynamical model (with $\gamma_{\rm eff} = 4/3$ corresponding to a compression factor of 7) yields a significantly higher explosion energy of $E=(2.76\pm0.66)\times10^{51}\, \text{erg}$ for the same ejecta mass. Apart from the unrealistic parameter values, we find that the dynamically-derived density for the $\gamma = 4/3$ case is inconsistent with  \textit{Spitzer} measurements of the post-shock density values from \citet{williams2011}. In contrast the two density measurements are in excellent  agreement for the $\gamma = 5/3$ case. We use these arguments to reject the $\gamma_{\rm eff} = 4/3$ model as invalid, providing further evidence that cosmic ray acceleration at the FS of \remnant\ is not efficient.

Our hydro-based MCMC analysis offers a robust and efficient pipeline to study SNRs and infer properties of the parent explosion from the kinematic information on the fluid discontinuities. In our review of the literature, we found several studies that use observational evidence from \remnosp's shock-heated gas in the context of hydro models to infer key explosion properties like explosion energy and mass. It is our strong opinion that such studies, complicated as they are by additional astrophysical effects such as electron-ion temperature equilibration, non-equilibrium ionization, potential heavy element radiative cooling, radial composition gradients in the ejecta, and so on, need to start with dynamical results obtained from clean shock kinematics. As we show here, kinematic data can significantly reduce the phase space of allowed solutions, reducing the number of gas-based models needed to be run. Additional kinematic constraints, such as improved RS locations and expansion speeds as well as deceleration rates, are potentially measurable with future observations. The ultimate goal is to obtain sufficient kinematic constraints from shocks to be able to constrain the ambient medium density, the remnant's age, explosion energy, mass, the effective equation of state at the FS, and possibly even the ejecta profile. This will likely require the inclusion of the wealth of information contained in the hot gas emission, but the kinematics of the shocks should lead the way.

\begin{acknowledgments}

P.A.\ thanks Peter Doze for helpful discussions about the science, and friends and family for encouragement and useful comments on the manuscript. J.P.H.\ acknowledges useful conversations with Ken Shen and thanks Zoe Rosenberg for help with validation using analytic models. We acknowledge support for this project from  the original HST grant in 2006 with grant number HST-GO-11015.001-A (PI: JPH), the second epoch HST program in 2016 with grant number HST-GO-14733.001-A  (PI: LH), NASA grant number NNX15AK71G (PI: JPH) and \chandra \ theory grant number TM0-21005X (PI: PA). A portion of this work was carried out at Los Alamos National Laboratory (LA-UR-21-32466), New Mexico. All of the data presented in this paper were obtained from the Mikulski Archive for Space Telescopes (MAST). STScI is operated by the Association of Universities for Research in Astronomy, Inc., under NASA contract NAS5-26555. All the \HST data used in this paper can be found at \href{http://dx.doi.org/10.17909/jka6-gg79}{DOI:10.17909/jka6-gg79}. This work has made use of data from the European Space Agency 
(ESA) mission {\it Gaia} (\url{https://www.cosmos.esa.int/gaia}), processed by the {\it Gaia} Data Processing 
and Analysis Consortium (DPAC, \url{https://www.cosmos.esa.int/web/gaia/dpac/consortium}). Funding for the DPAC 
has been provided by national institutions, in particular, the institutions participating in the {\it Gaia} 
Multilateral Agreement. This research made use of matplotlib: a Python library for publication-quality graphics \citep{Hunter:2007},Astropy: a community-developed core Python package for Astronomy \citep{astropy:2013,astropy:2018}, and NumPy \citep{harris2020array}. 
\end{acknowledgments}

\bibliographystyle{aasjournal}                   

\bibliography{super}

\end{document}